\documentclass[10pt]{article}
\usepackage{amsmath,amssymb,latexsym,epsfig,subfigure,multirow}
\newtheorem{lem}{Lemma}[section]

\newtheorem{pro}[lem]{Property}
\newtheorem{thm}[lem]{Theorem}

\newtheorem{definition}[lem]{Definition}
\newtheorem{remark}{Remark}

\newcommand\pf{{\noindent\bf Proof.}~}
\newcommand\qed{\hfill\framebox[2.5mm]{}}
\newcommand{\vsp}{\vspace{0.125in}}

\newcommand{\Comment}[1]{}

\begin{document}

\title{Modifications on Character Sequences and Construction of Large Even Length Binary Sequences}

\author{Tingyao Xiong\thanks{Partial support provided by the National
    Science Foundation} \\ {\tt xiongtin@msu.edu} \\and\\ Jonathan
  I. Hall\thanks{Partial support provided by the Institute for Quantum
    Sciences at Michigan State University and the National Science
    Foundation} \\ {\tt jhall@math.msu.edu} }

\date{}
\maketitle

{\begin{abstract} \textbf{It has been noticed that all the known binary sequences having the asymptotic merit factor $\ge 6$ are the modifications to the real primitive characters. In this paper, we give a new modification of the character sequences at length $N=p_1p_2\dots p_r$, where $p_i$'s are distinct odd primes and $r$ is finite. Based on these new modifications, for $N=p_1p_2\dots p_r$ with $p_i$'s distinct odd primes, we can construct a binary sequence of length $2N$ with asymptotic merit factor $6.0$.}
\end{abstract}}


\section{Introduction}

Let  $x=(x_0,x_1,\dots,x_{N-1})$ and $y=(y_0,y_1,\dots,y_{N-1})$ (not necessarily binary) be sequences of {\em length $N$}. The \itshape
aperiodic crosscorrelation \upshape function between $x$ and $y$ at shift $i$ is defined to be
\begin{equation}\label{eq:aperiodic between x and y}
A_{x,y}(i)=\sum^{N-i-1}_{j=0}x_{j}y_{j+i}\,, \quad 1\leq i\leq N-1\;.
\end{equation}

When $x=y$, denote
\begin{equation}\label{eq:aperiodic}
A_x(i)=A_{x,x}(i)=\sum^{N-i-1}_{j=0}x_{j}x_{j+i}\,, \quad 1\leq i\leq N-1\;,
\end{equation}
the \itshape aperiodic autocorrelation \upshape function of $x$ at shift $i$.

The \itshape periodic crosscorrelation \upshape function between $x$ and $y$ at shift $i$ is defined to be
\begin{equation}\label{eq:Periodic between x and y}
P_{x,y}(i)=\sum^{N-1}_{j=0}{x_{j}\,y_{j+i}},\;\;\;0\leq i\leq N-1\;,
\end{equation}
where all the subscripts are taken modulo $N$. Similarly, when $x=y$, put
\begin{equation}\label{eq:Periodic}
P_{x}(i)=P_{x,x}(i)=\sum^{N-1}_{j=0}{x_{j}\,x_{j+i}},\;\;\;0\leq i\leq N-1\;,
\end{equation}
the \itshape periodic autocorrelation \upshape function of $x$ at shift $i$ where all the subscripts are taken modulo $N$.

If the sequence $x$ is binary, which means that all the $x_j$'s are $+1$ or $-1$, the \itshape merit factor\upshape\; of the sequence $x$,
introduced by Golay \cite{Golay} in 1977, is defined as
\begin{equation}\label{eq:Merit Factor}
F_x=\frac{N^2}{2\sum^{N-1}_{i=1}A_x^2(i)}\,.
\end{equation}

Moreover, for a family of sequences
\[
S=\{x^1, x^2, \dots,x^n,\dots\}\;,
\]
where for each $i\geq 1$, $x^i$ is a binary sequence of increasing length $N_i$, if the limit of $F_{x^i}$ exists as $i$ approaches the infinity,
we call
\[
F=\lim_{i\rightarrow\infty}F_{x^i}\;,
\]
the asymptotic merit factor (ASF) of the sequence family $S$.

Since Golay firstly proposed the concept, the Merit Factor problems have attracted high research passion from mathematicians and engineers despite of the challenge. It is noticeable that all of the sequences with high asymptotic merit factor are derived from the primitive real characters. So first of all, we introduce the real primitive characters.
\begin{definition}\label{chiform}
Given an odd prime $p$,  the  real primitive character modulo $p$ is defined as
\[
\chi_p(j)=\left\{
\begin{array}{rl}
+1&,\; p\nmid j \text{ and }j \text{ is a square modulo } p,\\
-1&,\; p\nmid j \text{ and }j \text{ is a not square modulo } p,\\
0&,\; p\;|\;j
\end{array}
\right.
\]
More generally, for an odd number $N$, where $N= p_1p_2\dots p_r$ with $p_1<p_2<\dots <p_r$ distinct odd primes, the real primitive characters modulo $N$ take the form as
\begin{equation}\label{eq:chiform for N}
\chi_{N}(j)=\chi_{P_1}(j)\chi_{P_2}(j)\dots\chi_{P_r}(j)
\end{equation}
\end{definition}
Based on the character sequences, we define Legendre sequences and Jacobi sequences which are binary.
\begin{definition}\label{Legendre and Jacobi sequence}
For $p$ an odd prime, a Legendre sequence $\alpha=(\alpha_0,\alpha_1,\dots,\alpha_{p-1})$ of length $p$ is defined by the Legendre symbols
\begin{equation}\label{eq:Legendre symbol}
\alpha_j=\left(\frac{j}{p}\right)=\left\{\begin{array}{cl}
 1,&\text{ if }j=0\,;\\
\chi_p(j), & \text{ if } 1\leq j\leq p-1
\end{array}
\right.
\end{equation}
Generally, for $N=p_1p_2\dots p_r$, where $p_1<p_2<\dots p_r$ are distinct odd primes. A Jacobi sequence $J=(J_1,J_2,\dots,J_{N-1})$ is defined as
\begin{equation}\label{Jacobi sequence}
J_j=\left(\frac{j}{p_1}\right)\cdot\left(\frac{j}{p_2}\right)\dots\cdot\left(\frac{j}{p_r}\right)
\end{equation}
where $\left(\cdot\right)$ is the Legendre symbol as defined in expression (\ref{eq:Legendre symbol}).
\end{definition}
It is clear that the Legendre sequences, Jacobi and Modified Jacobi sequences just put new definitions at the i-th position where $(i,N) > 1$. Readers can find more discussion about the differences and similarity among Legendre sequences, Jacobi or modified Jacobi sequences and character sequences in \cite{TH2}.

Given a sequence $x=(x_0, x_1, \dots, x_{N-1})$ of length $N$, the periodic rotation
\[
x^r=(y_0,y_1,\dots,y_{N-1})
\]
of $x$ by a fraction $r$ is defined as
\begin{equation}\label{rotation form}
y_j=x_{j+\lfloor Nr\rfloor \pmod{N}},\quad 0\leq j<N.
\end{equation}

We define function
\begin{equation}\label{common form for amf}
F_r=\frac{2}{3}-4|r|+8r^2,\quad \text{ for real }r \text{ with }|r|\leq \frac{1}{2}.
\end{equation}

After 1980's, mathematicians have obtained a series of important results about the upper bound of asymptotic merit factors of binary sequences. In Table \ref{summary of known results}, we list all of known families of sequences with asymptotic merit factor form $\frac{1}{F_r}$. Note that for all the sequences listed in Table \ref{summary of known results}, $6.0$, the best theoretical proven value for the asymptotic merit factor, occurs at the rotation fraction $r=\frac{1}{4}$.
\begin{table}[h]\label{summary of known results}
\centering
\begin{tabular}[h]{ | c | c | c |c|c| }\hline
		Families of sequences & Length&Condition&Source\\[2.0ex]\hline
		Legendre sequence &$N=p_1$&$p_1\rightarrow \infty$&\cite{Hoholdt}\\[2.0ex]\hline
		Modified Jacobi sequence &$N=p_1p_2$ &$\frac{N^{\frac{1}{3}}}{p_1}\rightarrow 0$& \cite{Jensen} ,\cite{Borwein},\cite{Jedwab10},\cite{TH2}  \\[2.0ex]\hline
        Character sequences&$N=p_1p_2\dots p_k$&$\frac{(\log{N})^2}{p_1}\rightarrow 0$ &\cite{Borwein},\cite{Jedwab10}\\[2.0ex]\hline
\end{tabular}
\caption{ Summary of sequence families with high asymptotic merit factor values. }
\end{table}

Specifically, at length $N=p_1p_2$ with $p_1<p_2$ distinct odd primes, the authors and Jedwab and Schmidt have proved independently that any binary completion of the character sequence has the same asymptotic merit factor form $\frac{1}{F_r}$ under some restriction on $p_1$ and $p_2$ values. While Jedwab and Schmidt give a better condition on $p_1$ and $p_2$ values as shown in Table \ref{summary of known results}.
Therefore, we have
\[
 limsup_{n\rightarrow\infty}F_n\geq6.0
\]
where $F_n$ denotes the maximum value of the merit factor of all the binary sequences of length $n$. It has been observed \cite{Borwein6.34} that $limsup_{n\rightarrow\infty}F_n\geq6.34$ by appending a small fraction of a rotated Legendre sequence at an optimal ratio to the end of the rotated sequence itself.

In 2008, inspired by Parker's work, a doubling technique has been used to construct even length sequences with high asymptotic merit factor $6.0$ \cite{TH},\cite{Yu and Gong}, and the merit factor values for all the rotations of these even length sequences are computed in \cite{Schmidt}.

It is clear now that, without exception,  all the known binary sequences with high asymptotic merit factor $\geq 6.0$ are derived from the real primitive character sequences. Therefore, the real primitive character sequence $\chi_N$ has become (and seems will continue to be at least in the near future) a good candidate to generate sequences with high asymptotic merit factor.

Now suppose we start with the character sequence $\chi_N$, where $N=p_1p_2\dots p_r$, $p_i$'s are distinct odd primes, and $r$ is finite. If we want to construct a new sequence $z$ with high asymptotic merit factor based on $\chi_N$, we need to give the values to the $j$- positions such that $(j,N)>1$. Note that the number of  $j$'s such that $(j,N)>1= O(N^{\frac{r-1}{r}})$. Such a large amount  obviously deserves very careful construction. The main goal of this paper is giving a new construction on these positions where $N=p_1p_2\dots p_r$ with $r$ finite. Based on these new constructions, we could apply the doubling technique shown in \cite{TH} on $z$ successfully to get a new sequence $z'$ of length $2N$ with the best proven asymptotic merit factor $6.0$.
\section{Construction}\label{sec-const}
In the rest of the paper,  without confusion, we use notations $(i,N)$ or $i_N$ to represent $gcd(i,N)$. For both $A$ and $B$ positive, $A\ll B$ means that there exists a constant $k$ independent of $A$ and $B$, such that $|A|<kB$. And we will use the following notations heavily,
\begin{definition}\label{distinct prime}
For $n$ a positive integer, write $n=\prod_{i=1}^{r}p_i^{\alpha_i}$, where $p_i$'s are distinct primes. We define $\omega(n)=r$ to be the number of distinct prime divisors of $n$.
\end{definition}
\begin{definition}\label{sum of primes}
Let $n$ be a positive integer, and $f(x)$ be a function. Define
\[
\sum^n_{x=1} \mathtt{'} f(x)=\sum^n_{\stackrel{x=1}{(x,n)=1}}f(x)
\]
For example,
\[
\sum^4_{x=1} \mathtt{'} x^2=1^2+3^2=10.
\]
\end{definition}

Recall that in \cite{TH}, we gave the definition of a binary sequence to be symmetric or antisymmetric. For the convenience, we repeat the definition here:
\begin{definition}
For $N$ is odd, a sequence $\alpha=(\alpha_0,\alpha_1,\ldots,\alpha_{N-1})$ of length $N$ is {\em symmetric} if
$\alpha_i=\alpha_{N-i}$, for $1\leq i \leq N-1$, and {\em antisymmetric} if $\alpha_i=-\alpha_{N-i}$, for $1\leq i \leq N-1$.
\end{definition}

One concrete example of a symmetric or antisymmetric sequence is as in the following Lemma (\cite{TH2}, Lemma 3.5):
\begin{lem}\label{symmetricity of chi}
Let the character sequence $\chi_N$ be as defined in expression (\ref{eq:chiform for N}). Then $\chi_N$ is symmetric if $N\equiv 1\pmod{4}$, and antisymmetric if $N\equiv 3\pmod{4}$.
\end{lem}
\qed

Given a symmetric or antisymmetric sequence, a simple technique of exchanging the symmetric property is shown as following: (\cite{TH2}, Property 3.6)
\begin{pro}\label{symmetricity and antisymmetricity exchange}
Suppose $N$ is odd. For the sequence $\alpha=(\alpha_0,\alpha_1,\ldots,\alpha_{N-1})$ of length $N$, let the sequence $\beta=(\beta_0,\beta_1,\ldots,\beta_{N-1})$ with $\beta_j=(-1)^j\alpha_j$. If $\alpha$ is symmetric, then $\beta$ is antisymmetric,  while if $\alpha$ is antisymmetric, then $\beta$ is symmetric.
\end{pro}
\qed

To simplify the notations, from now on, we define the triple-valued sequence $U$ of length $N$ to be the character sequence
\begin{equation}\label{base sequence}
U_j=\left\{
\begin{array}{cl}\chi_N(j)&,\;j=1,\dots,N-1\;;\\
1&,\;j=0\;.
\end{array}
\right.
\end{equation}
%\begin{equation}\label{base sequence}
%U_j=\chi_N(j),\;\;\;\;\;\;\;\;\text{ where }0\leq j\leq N-1
%\end{equation}
By Lemma \ref{symmetricity of chi}, it is easy to see that sequence $U$ is symmetric if $N\equiv 1\pmod{4}$, and antisymmetric if $N\equiv 3\pmod{4}$.

Before we give the construction to the sequences which will be discussed throughout the whole paper, we study a concrete example.
\newline

\textbf{Example 1}\hspace{1cm}Suppose $N=3\times 5\times 7=105$, sequence $U$ of length $105$ is as defined in expression (\ref{base sequence}), and Jacobi sequence $J$ of length $105$ is as shown in expression (\ref{Jacobi sequence}). Since we are only interested in the positions $j$ with $(j,N)>1$. So we only list some positions $j$ with $(j,N)>1$ of sequences $U$ and $J$ in the Table \ref{U and J for N=105}:
\begin{table}[h]\label{U and J for N=105}
\caption{Comparison between character sequence and Jacobi Sequence at length $N=105$}
\vspace{0.4cm}
\begin{tabular}{c c c c c c c c c c c}
position $j$	&$3$&$5$&$6$&$7$&$9$&$10$&$12$&$14$&$15$&$\ldots$\\
$U_j$           &$0$&$0$&$0$&$0$&$0$&$0$&$0$&$0$&$0$&$\ldots$\\
$J_j$           &$1$&$1$&$-1$&$1$&$1$&$-1$&$1$&$1$&$1$&$\ldots$\\ &$\stackrel{\uparrow}{\chi_{35}(1)}$&$\stackrel{\uparrow}{\chi_{21}(1)}$&$\stackrel{\uparrow}{\chi_{35}(2)}$&$\stackrel{\uparrow}{\chi_{15}(1)}$&$\stackrel{\uparrow}{\chi_{35}(3)}$&$\stackrel{\uparrow}{\chi_{21}(2)}$&$\stackrel{\uparrow}{\chi_{35}(4)}$&$\stackrel{\uparrow}{\chi_{15}(2)}$&$\stackrel{\uparrow}{\chi_{7}(1)}$&$\ldots$
\end{tabular}
\end{table}
\begin{table}[h]\label{comparation2}
\begin{tabular}{ c c c c c c c c c c c }
position $j$	&$\ldots$&$90$&$91$&$93$&$95$&$96$&$98$&$99$&$100$&$102$\\
$U_j$           &$\ldots$&$0$&$0$&$0$&$0$&$0$&$0$&$0$&$0$&$0$\\
$J_j$           &$\ldots$&$-1$&$-1$&$-1$&$-1$&$-1$&$-1$&$1$&$1$&$-1$\\
&$\ldots$&$\stackrel{\uparrow}{\chi_{7}(6)}$&$\stackrel{\uparrow}{\chi_{15}(13)}$&$\stackrel{\uparrow}{\chi_{35}(31)}$&$\stackrel{\uparrow}{\chi_{21}(19)}$&$\stackrel{\uparrow}{\chi_{35}(32)}$&$\stackrel{\uparrow}{\chi_{15}(14)}$&$\stackrel{\uparrow}{\chi_{35}(33)}$&$\stackrel{\uparrow}{\chi_{21}(20)}$&$\stackrel{\uparrow}{\chi_{35}(34)}$
\end{tabular}
\end{table}

In Example 1, $U$ is symmetric because $105\equiv 1\pmod{4}$. But Jacobi sequence $J$ is neither symmetric nor antisymmetric because as shown above, some positions give subsequences which are antisymmetric. In Table \ref{antisymmetric subsequences}, we list the positions which give antisymmetric subsequences within a Jacobi sequence of length $N=105$.
\begin{table}[h]
\caption{antisymmetric subsequences inside Jacobi sequence $J$ at length $N=105$}\label{antisymmetric subsequences}
\vspace{0.3cm}
\begin{tabular}{c c}
\hspace{1cm}positions & \hspace{2cm}corresponding subsequence\\[1ex]
\hspace{1cm}$0$,\ \ \ $35$,\ \ \ $70$\;&\hspace{2cm} $(1$,\ \ \ $\chi_{3}(1)$,\ \ \ $\chi_{3}(2)) $\\[1ex]
\hspace{1cm}$0$,\ \ \ $15$,\ \ \ $30$,\ \ \ $\dots$,\ \ \ $75$,\ \ \ $90$&\hspace{2cm} $(1$,\ \ \ $\chi_{7}(1)$,\ \ \ $\chi_{7}(2)$,\ \ \ $\ldots$, \ \ \ $\chi_{7}(5)$,\ \ \ $\chi_{7}(6))$\\[1ex]
\hspace{1cm}$0$,\ \ \ $7$,\ \ \ $14$,\ \ \ $\dots$,\ \ \ $91$,\ \ \ $98$& \hspace{2cm} $(1$,\ \ \ $\chi_{15}(1)$,\ \ \ $\chi_{15}(2)$, $\ldots$,\ \ \ $\chi_{15}(13)$,\ \ \ $\chi_{15}(14) )$\\[1ex]
\hspace{1cm}$0$,\ \ \ $3$,\ \ \ $6$,\ \ \ \dots,\ \ \ $99$,\ \ \ $102$&\hspace{2cm} $(1$\ \ \, $\chi_{35}(1)$,\ \ \ $\chi_{35}(2)$, $\ldots$,\ \ \ $\chi_{35}(33)$,\ \ \ $\chi_{35}(34))$\\
\end{tabular}
\end{table}
The subsequences listed in Table \ref{antisymmetric subsequences} are all antisymmetric since $d\equiv 3\pmod{4}$, where $d\in \{3,7,15,35\}$. This is consistent to Lemma \ref{symmetricity of chi}. In the following constructions, we give new definitions on positions $j$ in Table \ref{new construction}. Then the new sequences have the same symmetric property as sequence $U$.

\begin{table}[h]
\caption{modified subsequences with the same symmetric property as sequence $U$.}\label{new construction}
\vspace{0.3cm}
\begin{tabular}{c c}
positions &\hspace{1cm} corresponding subsequence\\[1.5ex]
$0$,\ \ \ $35$,\ \ \ $70$\;& \hspace{1cm}$(1$,\ \ \ $(-1)^1\chi_{3}(1)$,\ \ \ $(-1)^2\chi_{3}(2)) $\\[1.5ex]
$0$,\ \ $15$,\ \ $30$,\ \ $\dots$,\ \ $75$,\ \ $90$&\hspace{1cm} $(1$,\ \ \ $(-1)^1\chi_{7}(1)$,\ \ \ $(-1)^2\chi_{7}(2)$,\ \ \ $\ldots$, \ \ \ $(-1)^5\chi_{7}(5)$,\ \ \ $(-1)^6\chi_{7}(6))$\\[1.5ex]
$0$,\ \ $7$,\ \ $14$,\ \ $\dots$,\ \ $91$,\ \ $98$&\hspace{1cm}  $(1$,\ \ \ $(-1)^1\chi_{15}(1)$,\ \ \ $(-1)^2\chi_{15}(2)$, $\ldots$,\ \ \ $(-1)^{13}\chi_{15}(13)$,\ \ \ $(-1)^{14}\chi_{15}(14) )$\\[1.5ex]
$0$,\ \ $3$,\ \ $6$,\ \ \dots,\ \ $99$,\ \ $102$&\hspace{1cm} $(1$\ \ \, $(-1)^1\chi_{35}(1)$,\ \ \ $(-1)^2\chi_{35}(2)$, $\ldots$,\ \ \ $(-1)^{33}\chi_{35}(33)$,\ \ \ $(-1)^{34}\chi_{35}(34))$\\
\end{tabular}
\end{table}

\vspace{0.3cm}

In Table \ref{new construction}, we change the subsequence $(1$, $\chi_{d}(1)$, $\chi_{d}(2)$, $\ldots$, $\chi_{d}(d-2)$, $\chi_{d}(d-1))$ into $(1$, $(-1)^1\chi_{d}(1)$, $(-1)^2\chi_{d}(2)$, $\ldots$ , $(-1)^{d-2}\chi_{d}(d-2)$, $(-1)^{d-1}\chi_{d}(d-1))$, where $d\in \{3,7,15,35\}$. From Property \ref{symmetricity and antisymmetricity exchange}, we know that all of the new subsequences become symmetric after the modification. As a result, the new sequence is symmetric, or it has the same symmetric property as sequence $U$.

We generalize the idea of Example 1 as in the following definitions.
\begin{definition}\label{phi}
Let $N=p_1p_2\ldots p_r$, where $p_i$'s are distinct odd primes and $r\geq 2$, for $1\leq j\leq N-1$, define
\begin{equation}\label{sequence phi}
v_j=\left\{
\begin{array}{cl}
\chi_{N/d}(j/d)&,\;if\ (j,N)=d>1\ and\ N\equiv N/d\; \pmod{4}\\
&\\
(-1)^{j/d}\chi_{N/d}(j/d)&,\;if\ (j,N)=d>1\ and\ N\not\equiv N/d\; \pmod{4}\\
&\\
0&,\;otherwise\\
\end{array}
\right.
\end{equation}
and
\begin{equation}\label{sequence Z}
z_j=\left\{
\begin{array}{cl}
1&,\;if\ j=0;\\
v_j&,\;if\ (j,N)>1;\\
U_j&,\;otherwise
\end{array}
\right.
\end{equation}
\end{definition}
where $U$ is character sequences defined in expression (\ref{base sequence}).
\vsp
\newline

The following Lemma shows that the sequence $z$ defined in Definition \ref{phi} has the same symmetric property as the character sequence $U$.
\begin{lem}\label{sym or antisym}
Suppose $N=p_1p_2\dots p_r$, where $p_i$'s are distinct odd primes. And the binary sequence $z$ of length $N$ is as defined in Definition \ref{phi}. Then $z$ is symmetric if $N\equiv 1 \pmod{4}$, and $z$ is antisymmetric if $N\equiv 3 \pmod{4}$.
\end{lem}
\pf
From Definition \ref{phi}, it is sufficient to show that $v$ as defined in Definition \ref{phi} above has the same symmetric property as the character sequence $U$.
\newline
By the definition of sequence $v$, $v_j=0$ if $(j,N)=1$. Then $v_{N-j}=0$ if $(j,N)=1$. So we only need to consider $v_j$ values when $(j,N)=d>1$.
\newline
Furthermore, $v_j=\chi_{N/d}(j/d)$ if $N/d\equiv N \pmod{4}$, $v_j=(-1)^{j/d}\chi_{N/d}(j/d)$ if $N/d\not\equiv N \pmod{4}$.
\begin{enumerate}
  \item[(1)] When $N/d\equiv N \pmod{4}$, Lemma \ref{symmetricity of chi} shows that $\chi_{N/d}$ has the same symmetric property as character sequence $U$. Or $\chi_{N/d}$ is symmetric when $N/d\equiv N\equiv 1 \pmod{4}$, and $\chi_{N/d}$ is antisymmetric when $N/d\equiv N\equiv 3 \pmod{4}$.
  \item[(2)] When $N/d\not \equiv N\pmod{4}$, again by Lemma \ref{symmetricity of chi}, $\chi_{N/d}$ has the opposite symmetric property to character sequence $U$. While Property \ref{symmetricity and antisymmetricity exchange} shows that $\{(-1)^{j/d}\chi_{N/d}(j/d)\;|\;j=0,1,\dots,N/d-1\}$ alters the symmetric property of sequence $\chi_{N/d}$.
\end{enumerate}
We combine the discussion above, and we can claim that sequence $v$ has the same symmetric property as the character sequence $U$. In other words,
$z$ is symmetric if $N\equiv 1 \pmod{4}$, and $z$ is antisymmetric if $N\equiv 3 \pmod{4}$.
\qed
\vsp
\newline
Next we review some definitions from \cite{TH}.
\begin{definition}
Given two binary sequences $x=(x_0,x_1,\ldots,x_{N-1})$ and
$e=(e_0,e_1,\ldots,e_{N-1})$, we define a new sequence $\{x,x\}=(x_0,x_1,\ldots,x_{N-1},x_0,x_1,\ldots,x_{N-1})$ of length $2N$. And we define the product sequence $b=x\ast e$ by $b_i=x_i e_i$, for $i=0,1,...,N-1$.
\end{definition}
\begin{definition}\label{defn-seq}
For $\delta=0,1$, let the four sequences $\pm e^{(\delta)}$ be
given by
\begin{equation}
\label{eq:Epsi}
e_j^{(\delta)}=(-1)^{\left(\stackrel{j+\delta}{2}\right)}
\end{equation}
\end{definition}
The main goal of this paper is to prove the following theorem.
\begin{thm}(main theorem)\label{thm2}
For any positive integer $r\geq 2$, suppose $N=p_1p_2\ldots p_r$, where $p_1<p_2<\ldots p_r$ are distinct odd primes. Let $z$ be the binary sequences defined in Definitions \ref{phi}, then
\begin{itemize}
  \item [(1)]Let $F$ be the asymptotic merit factor of $z$, $f$ be the offset fraction. Then we have
\[
1/F=2/3-4|f|+8f^2, \;\;\;\;\;\;\;|f|\leq 1/2\,,\;given
\]
\begin{equation}\label{epsilon and p_1}
\frac{N^\epsilon}{p_1}\rightarrow \infty {\;\text{for any }} \epsilon>0 {\text{ small enough as }}N\rightarrow \infty.
\end{equation}
  \item [(2)]Let the sequence $e$ of length $2N$ be one of the four sequences $\pm e^{(\delta)}$ from the Definition \ref{defn-seq}. The new sequence $b=\{z,z\}*e$ of length $2N$ has asymptotic merit factor $6.0$ given (\ref{epsilon and p_1}) is satisfied.
\end{itemize}
\end{thm}
The key step of proving Theorem \ref{thm2} is to show that $\sum_{i=1}^{N-1}P_z^2(i)\sim o(N^2)$ when $N$ is large. So we will estimate the periodic autocorrelations of sequence $z$ in the following section.

\section{Periodic Autocorrelations of Sequences z}
In this section, we will prove that when $N$ is large, $\sum_{i=1}^{N-1}P_z^2(i)\sim o(N^2)$ given condition (\ref{epsilon and p_1}) is held. We will prove this result in three steps.
\begin{enumerate}
  \item[(a)] In section \ref{section 3.1}, we will prove that $\sum_{i=1}^{N-1}P_U^2(i)\sim o(N^2)$.
  \item[(b)] In section \ref{section 3.2}, we will prove that $\sum_{i=1}^{N-1}P_v^2(i)\sim o(N^2)$
  \item[(c)] Finally, in in section \ref{section 3.3}, we will prove that $\sum_{i=1}^{N-1}P_z^2(i)\sim o(N^2)$
\end{enumerate}
given condition (\ref{epsilon and p_1}) is satisfied.
\subsection{Upper Bound for $\sum_{i=1}^{N-1}P_U^2(i)$.}\label{section 3.1}
Firstly, let's review some simple properties from number  theory. A well known result about the primitive real characters modulo prime $p$ is as following
\begin{lem}\label{chi}
Suppose $\chi_p$ is as defined in Definition \ref{chiform}, then
\[
\sum_{n=0}^{p-1}\chi_p(n)\chi_p(n-k)=\left\{\begin{array}{rl}
p-1 &,\; if \ p|k;\\
-1&,\;otherwise\\
\end{array}
\right.
\]
\end{lem}
\pf
Readers can find the proof to Lemma \ref{chi} in many references, for instance, Lemma 2 in \cite{Brian}.
\qed

\begin{definition}\label{divisor function}
For an integer $n$, the divisor function $d(n)$, is defined to be the number of positive divisors of $n$, or
\[
d(n)=\sum_{0<d|n}1
\]
\end{definition}
\begin{definition}\label{DFT}
Let $\xi_N^j=e^{\frac{2\pi j}{N}i}$, $u=(u_0,u_1,\dots,u_{N-1})$ be a binary sequence of length $N$. The Discrete Fourier Transform (D.F.T.) of sequence $u$ is defined as
\[
u(\xi_N^j)=\sum_{i=0}^{N-1}u_i\cdot(\xi_N^j)^i.
\]
\end{definition}
Given a positive integer $N$, recall that Euler function $\phi(N)$ is defined as the number of $i$ such that $(i,N)=1$. Then we have
\begin{lem}\label{Euler}
Let $N=p_1p_2\dots p_r$, where $p_1<p_2<\dots <p_r$' are distinct odd primes. Then
\[
N-\phi(N)< r\times \frac{N}{p_1}
\]
where $\phi(N)=|\{i|(i,N)=1\}|$ is the Euler function of $N$.
\end{lem}
\pf
\[
N-\phi(N)<\sum_{i=1}^r(N/p_i-1)< r\times\frac{N}{p_1}
\]
\qed
\begin{lem}\label{periodic correlation of sequence product}
Let $y^1,y^2,\dots y^r$ be $r$ sequences (not necessarily binary) of length $N_1,N_2,\dots N_r$ respectively, such that $(N_i,N_j)=1$ for any $1\leq i,j\leq r$. Let $N=N_1\times N_2\dots \times N_r$, define a new sequence $u=y^1\ast y^2\ast \dots y^r$ of length $N$ via
\[
u_m=\prod_{s=1}^r y^s_{m}, \text{ where }0\leq m\leq N-1.
\]
Then the periodic autocorrelations of $u$ is
\[
P_u(n)=\prod^r_{s=1}P_{y^s}(n), \text{ where }0\leq n\leq N-1.
\]
Let $\xi_N^j=e^{\frac{2\pi j}{N}i}$, let $u(\xi_N^j)$ be the D.F.T. as defined in Definition \ref{DFT}. Then there exist integers $s_1,s_2,\dots, s_r$ with $(s_i,N_i)=1$ for $1\leq i\leq r$, then
\[
u(\xi_N^j)=\prod_{i=1}^r y^i(\xi_{N_i}^{js_i})
\]
\end{lem}
\pf
We will prove the two results simultaneously by the induction on $r$. When $r=1$, the result is trivial.
Now suppose  Lemma \ref{periodic correlation of sequence product} holds for $r-1$, where $r\geq 2$ then for $r$, suppose $y^1,y^2,\dots y^{r-1},y^r$ is a series of sequences, where for each $i$, sequence $y^i$ has length $N_i$, and $(N_i,N_j)=1$ for any $1\leq i< j\leq r$. Now denote $N'=N_1\times N_2\dots \times N_{r-1}$,
$u_1=y^1\ast y^2\ast \dots y^{r-1}$, then $u=u_1\ast y^{r}$. By induction,
\[
P_u(n)=P_{u_1}(n)P_{y^{r}}(n)
\]
\[
u(\xi_N^j)=u_1(\xi_{N'}^{js'})\times y^{r}(\xi_{N_r}^{js_{r}})
\]
where $(s',N')=1$ and $(s_{r}, N_{r})=1$. Then by induction
\[
P_u(n)=P_{u_1}(n)P_{y^{r}}(n)=\prod^{r-1}_{i=1}P_{y^i}(n)\cdot P_{y^r}(n)=\prod^{r}_{i=1}P_{y^i}(n)
\]
\newline
On the other hand, by induction,
\[
u_1(\xi^{js'}_{N'})=\prod_{i=1}^{r-1}y^i(\xi^{js's'_i}_{N_i})
\]
where $(s'_i,N_i)=1$, for $i=1,2,\dots, r-1$. Since $(s',N')=1\Rightarrow (s',N_i)=1$, for $i=1,2,\dots, r-1$. Thus $ (s's'_i,N_i)=1$, we denote $s_i=s's'_i$, then
\[
u(\xi_N^j)=\prod_{i=1}^{r-1}y^i(\xi^{js_i}_{N_i})\cdot y^r(\xi^{js_r}_{N_r})=\prod_{i=1}^{r}y^i(\xi^{js_i}_{N_i})
\]
This finishes the proof of the Lemma.
\qed
\vsp
\newline
Now we consider a simple example of Lemma \ref{periodic correlation of sequence product}. Let sequence $U$ be as defined in form (\ref{base sequence}). If $r=2$, so $N=pq$, where $p$ and $q$  are different odd primes. Then from Lemma \ref{chi} and \ref{periodic correlation of sequence product}
\begin{equation}\label{periodic form for pq}
P_{U}(i)=\left\{\begin{array}{rl}
1-p &,\; if \ p\;|\;i\\
1-q &,\; if \ q\;|\;i\\
+1&,\;otherwise\\
\end{array}
\right.
,\;\; \text{ where }1\leq i\leq N-1.
\end{equation}
Generally, for  $r\geq 2$ is finite, $N=p_1p_2\dots p_r$, where $p_i$'s are distinct odd primes, we have the following upper estimate for the periodic autocorrelation for $U$ based on Lemma \ref{periodic correlation of sequence product}.
\begin{lem}\label{P for product}
Let $N=p_1p_2\dots p_r$, where $p_1<p_2<\dots <p_r$ are distinct odd primes, $r$ is finite. Let the sequence $U$ of entries $\{0,\pm1\}$ be as defined in form (\ref{base sequence}), then we have
\begin{enumerate}
\item[(a)] $|P_U(i)|\leq (i,N)$;
\item[(b)] $\sum_{i=1}^{N-1}P^2_U(i)\leq c\frac{N^2}{p_1}$, where $c$ is a constant only depends on $r$.
\end{enumerate}

\end{lem}
\pf
For part $(a)$, if  $(i,N)=1$, then $(i,p_j)=1$ for $j=1,2,\dots, r$. From Lemma \ref{chi} and \ref{periodic correlation of sequence product},
\[
|P_U(i)|=|\prod_{j=1}^{r}P_{\chi_{p_j}}(i)|=1=(i,N)
\]
Now if $(i,N)=N_1>1$, then $(i, N/N_1)=1$. Use the above result and Lemma \ref{periodic correlation of sequence product},
\[
|P_U(i)|=|P_{\chi_{N_1}}(0)\times P_{\chi_{N/N_1}}(i)|= |P_{\chi_{N_1}}(0)|\leq (i,N)
\]
So this finishes the proof of part (a).
\vsp
\newline

For part $(b)$, by Lemma \ref{periodic correlation of sequence product},
\begin{align}\nonumber
\sum_{i=1}^{N-1}P_U^2(i)&=\sum_{(i,N)=1}P_U^2(i)+\sum_{(i,N)>1}P_U^2(i)\\\nonumber
&\leq\sum_{(i,N)=1}\prod_{j=1}^rP^2_{\chi_{p_j}}(i)+\sum_{1<d|N}\sum^{\frac{N}{d}}_{s=1} \mathtt{'}P^2_{\chi_d}(sd)P^2_{\chi_{N/d}}(sd)\\\nonumber
&\leq \sum_{(i,N)=1}1+\sum_{1<d|N}d^2\cdot \frac{N}{d}\\\nonumber
&=\phi(N)+\sum_{1<d|N}N\cdot d\\\nonumber
&\leq c\frac{N^2}{p_1}\hspace{0.5cm}\text{ where $c$ only depends on }r
\end{align}
where $\phi(N)$ is the Euler function of $N$.\qed
\vsp
\newline
So we have proved that $\sum_{i=1}^{N-1}P_U^2(i)\sim O(\frac{N^2}{p_1})\sim o(N^2)$ given condition (\ref{epsilon and p_1}) is held.
\subsection{Upper Bound for $\sum_{i=1}^{N-1}P_v^2(i)$}\label{section 3.2}
Before we could give an upper bound of the periodic autocorrelations of sequence $v$, we still need several properties.
\vsp
\newline
Let $\xi_N^k=e^{\frac{2\pi ki}{N}}$, we have the following lemma
\begin{lem}\label{seperate e}
Suppose $N=N_1\times N_2\dots \times N_r$, where $(N_i,N_j)=1$, for any $1\leq i<j\leq r$, then for any integer $k$, there exist integers $k_1,k_2,\dots,k_r$, such that $(k_i,N_i)=1$, and
\[
\xi_N^k=\prod_{i=1}^r\xi_{N_i}^{kk_i}
\]
\end{lem}
\pf
We will prove the lemma by induction on $r$. When $r=1$, the result is obviously true if we choose $k_1=1$. Suppose the result is correct for $r-1$, where $r\geq 2$. Then for $r$, so $N=N_1\times N_2\dots \times N_{r-1}\times N_r$, where $(N_i,N_j)=1$, for any $1\leq i<j\leq r$. Denote $N'=N_1\times N_2\dots \times N_{r-1}$, so $(N',N_r)=1$. By induction, there exist integers $k'$ and $k_r$, with $(k', N')=1$, $(k_r, N_r)=1$, such that
\begin{equation}
\xi_N^k=\xi_{N'}^{kk'}\xi_{N_r}^{kk_r}
\end{equation}
by induction
\[
\xi_{N'}^{kk'}=\prod_{i=1}^{r-1}\xi_{N_i}^{kk's_i}
\]
where $(s_i,N_i)=1$, for $1\leq i\leq r-1$. Let $k_i=k's_i$, then $(k_i, N_i)=1$, for $1\leq i\leq r-1$. Then we have
\[
\xi_N^k=\prod_{i=1}^{r-1}\xi_{N_i}^{kk_i}\cdot \xi_{N_r}^{kk_r}=\prod_{i=1}^{r}\xi_{N_i}^{kk_i}
\]
\qed
\vsp
\begin{lem}\label{Weil for general}
Suppose $N=p_1p_2\dots p_r$, where $p_i$'s are distinct odd primes for $i=1,2,\dots,r$. Let $\chi_N$ be the primitive character mod $N$, $f(x)$ be a polynomial of degree $k$. If for each $p_a$, $1\leq a\leq r$, a factorization $f(x)=b(x-x_1)^{d_1}\dots (x-x_s)^{d_s}$ in $\overline{F}_{p_a}$, where $x_i\neq x_j$, for $i\neq j$ with
\[
(p_a-1,d_1,\dots,d_s)=1.
\]
Then
\[
|\sum_{u<n\leq u+t}\chi_N(f(n))|< 2k^rN^{\frac{1}{2}}\log{(N)}
\]
where $u$ and $t$ are integers and $0<t<N$.
\end{lem}
\pf
From the Lemma 3 in \cite{M-S} (Page 374), we know that for each $p_j$, $1\leq j\leq r$,
\begin{equation}\label{sch1}
|\sum_{x\in F_{p_j}}\chi_{p_j}(f(x))e^{\frac{2b\pi}{p_j}i}|\leq kp_j^{\frac{1}{2}}
\end{equation}
for any $b\in \mathbb{Z}$. At the same time, one form of the Erd\H{o}s-Tur\'{a}n inequality (\cite{M-S} Lemma 4, Page 375) is presented as following
\newline
If $m\in$ $\mathbb{N}$, the function $g(x)$: $\mathbb{Z}$ $\rightarrow$$\mathbb{C}$ is periodic with period $m$, and $u$ and $t$ are real numbers with $0\leq t<m$, then
\begin{equation}\label{E-T inequality}
|\sum_{u<n\leq u+t}g(n)|\leq \frac{t+1}{m}|\sum_{n=1}^{m}g(n)|+\sum_{1\leq |h|\leq m/2}|h|^{-1}|\sum_{n=1}^{m}g(n)e^{\frac{hn2\pi}{m}i}|
\end{equation}
Now apply equation (\ref{E-T inequality}) with $N$ and $\chi_N(f(n))$ in place of $m$ and $g(n)$ respectively, and use Lemma \ref{periodic correlation of sequence product} and equation (\ref{sch1})
\begin{align}\label{set up}
|\sum_{u<n\leq u+t}\chi_N(f(n))|& \leq \frac{t+1}{N}|\sum_{n=1}^{N}\chi_N(f(n))|+\sum_{1\leq |h|\leq N/2}|h|^{-1}|\sum_{n=1}^N\chi_N(f(n))e^{\frac{hn2\pi}{N}i}|\\\nonumber
&=\frac{t+1}{N}\prod_{j=1}^{r}|\sum_{n=1}^{p_j}\chi_{p_j}(f(n))|+\sum_{1\leq |h|\leq N/2}|h|^{-1}\prod_{j=1}^{r}|\sum_{n=1}^{p_j}\chi_{p_j}(f(n))e^{\frac{hk_jn2\pi}{p_j}i}|
\end{align}
where  $k_j$'s are integers such that $(k_j,p_j)=1$, for $1\leq j\leq r$.
\newline
From expression (\ref{sch1}), the first item of (\ref{set up}) satisfies
\[
\frac{t+1}{N}\prod_{j=1}^{r}|\sum_{n=1}^{p_j}\chi_{p_j}(f(n))|\leq \frac{t+1}{N}\prod_{j=1}^{r}kp_j^{\frac{1}{2}}\leq k^rN^{\frac{1}{2}}
\]
For the second item in (\ref{set up}), using (\ref{sch1}), we have
\[
\sum_{1\leq |h|\leq N/2}|h|^{-1}\prod_{j=1}^{r}|\sum_{n=1}^{p_j}\chi_{p_j}(f(n))e^{\frac{hk_jn2\pi}{p_j}i}|\leq \sum_{1\leq |h|\leq N/2}|h|^{-1}k^rN^{\frac{1}{2}}
 \]
Therefore we obtain
\begin{align*}
|\sum_{u<n\leq u+t}\chi_N(f(n))|&\leq \frac{t+1}{N}\prod_{j=1}^{r}|\sum_{n=1}^{p_j}\chi_{p_j}(f(n))|+\sum_{1\leq |h|\leq N/2}|h|^{-1}\prod_{j=1}^{r}|\sum_{n=1}^{p_j}\chi_{p_j}(f(n))e^{\frac{hk_jn2\pi}{p_j}i}|\\
&\leq k^rN^{\frac{1}{2}}+k^rN^{\frac{1}{2}}\sum_{1\leq |h|\leq N/2}|h|^{-1}\\
&< 2k^rN^{\frac{1}{2}}\log{(N)}
\end{align*}
which is the desired result we want to prove.
\qed
\begin{remark}\label{remark1}
In the hypothesis of Lemma \ref{Weil for general}, for each $p_j$, $1\leq j\leq r$, $f(x)$ can't be a perfect square over $\overline{F}_{p_j}$. As an application of Lemma \ref{Weil for general}, the following property gives a general estimate for all $f(x)$ of degree $2$.
\end{remark}
\begin{pro}\label{f(x) of order 2}
Suppose $N=p_1p_2\dots p_r$, where $p_i$'s are distinct odd primes and $r$ is finite. Let $\chi$ be the primitive character modulo $N$. Let $u$ and $t$ be integers such that $0\leq t<N$, then for any $1\leq k_1\neq k_2\leq N-1$, we have
\begin{enumerate}
  \item[1.] $|\sum_{u<n\leq u+t}\chi_N(n+k_1)\chi_N(n+k_2)|\ll max\{d, \sqrt{N/d}\; \log{(N/d)}\}$
  \item[2.] $|\sum_{u<n\leq u+t}(-1)^n\chi_N(n+k_1)\chi_N(n+k_2)|\ll max\{d, \sqrt{N/d}\;\log{(N/d)}\}$
\end{enumerate}
where $d=(k_2-k_1,N)$.
\end{pro}
\pf
We will prove part $1$ firstly. For $d=(k_2-k_1,N)$, write $N=ds$, thus $(k_2-k_1,s)=1$, then
\begin{align*}
|\sum_{u<n\leq u+t}\chi_N(n+k_1)\chi_N(n+k_2)|&=|\sum_{u+k_1<n\leq u+t+k_1}\chi_N(n)\chi_N(n+k_2-k_1)|\\
&=|\sum_{u'<n\leq u'+t}\chi_d(n)\chi_{s}(n)\chi_{d}(n+k_2-k_1)\chi_{s}(n+k_2-k_1)|\\
&=|\sum_{u'<n\leq u'+t}\chi_d^2(n)\chi_{s}(n)\chi_{s}(n+k_2-k_1)|\hspace{1.0cm}\text{ since }d\;|(k_2-k_1)\\
&=|\sum_{u'<n\leq u'+t}\chi_{s}(n)\chi_{s}(n+k_2-k_1)|
\end{align*}
where $u'=u+k_1$. Let $m=\lfloor\frac{t}{s}\rfloor=\lfloor\frac{td}{N}\rfloor\leq d$, since $t<N$.
\[
|\sum_{u'<n\leq u'+t}\chi_N(n)\chi_N(n+k_2-k_1)|\leq \sum_{j=1}^m|P_{\chi_s}(k_2-k_1)|+ |\sum_{u'+ms<n\leq u'+t}\chi_{s}(n)\chi_{s}(n+k_2-k_1)|
\]
Because $(k_2-k_1,s)=1\Rightarrow P_{\chi_s}(k_2-k_1)=-1$, we have
\begin{align*}
|\sum_{u'+ms<n\leq u'+t}\chi_{s}(n)\chi_{s}(n+k_2-k_1)|&\leq m+|\sum_{u'+ms<n\leq u'+t}\chi_{s}(n)\chi_{s}(n+k_2-k_1)|\\
&\leq 2\cdot max\{m,\; 2^{\omega+1}\sqrt{s}\;\log{(s)}\}\hspace{2.5cm}\text{by Lemma \ref{Weil for general}}\\
&\leq 2\cdot max\{d,\; 2^{\omega+1}\sqrt{N/d}\;\log{(N/d)}\}\hspace{1.5cm}\text{ since }m\leq d
\end{align*}
where $\omega=\omega(s)=\omega(N/d)$ is as defined in Definition \ref{distinct prime}. Therefore,
\[
|\sum_{u'+ms<n\leq u'+t}\chi_{s}(n)\chi_{s}(n+k_2-k_1)|\ll max\{d, \sqrt{N/d}\;\log{(N/d)}\}
\]
For part $2$.
\begin{align}\label{alter general}\nonumber
  |\sum_{u<n\leq u+t}(-1)^n\chi_N(n+k_1)\chi_N(n+k_2)|&\leq |\sum_{\lfloor\frac{u}{2}\rfloor<n\leq \lfloor\frac{u+t}{2}\rfloor}(-1)^{2n}\chi_N(2n+k_1)\chi_N(2n+k_2)\\\nonumber
  &+\sum_{\lfloor\frac{u}{2}\rfloor<n\leq \lfloor\frac{u+t}{2}\rfloor}(-1)^{2n-1}\chi_N(2n-1+k_1)\chi_N(2n-1+k_2)|+2\\\nonumber
  &\leq|\sum_{\lfloor\frac{u}{2}\rfloor<n\leq \lfloor\frac{u+t}{2}\rfloor}\chi_N(2n+k_1)\chi_N(2n+k_2)|\\ &+|\sum_{\lfloor\frac{u}{2}\rfloor<n\leq \lfloor\frac{u+t}{2}\rfloor}\chi_N(2n-1+k_1)\chi_N(2n-1+k_2)|+2
\end{align}
For the first item in expression (\ref{alter general}), similarly to the above calculation, we have
\begin{align}\label{alter general 1}\nonumber
|\sum_{\lfloor\frac{u}{2}\rfloor<n\leq \lfloor\frac{u+t}{2}\rfloor}\chi_N(2n+k_1)\chi_N(2n+k_2)|&=|\sum_{\lfloor\frac{u}{2}\rfloor<n\leq \lfloor\frac{u+t}{2}\rfloor}\chi_N(n+2^{-1}k_1)\chi_N(n+2^{-1}k_2)|\\
&\ll  max\{d, \sqrt{N/d}\; \log{(N/d)}\}
\end{align}
from part $1$.
\vsp
\newline
Similarly, from part $1$, the second item in expression (\ref{alter general}),
\begin{align}\label{alter general 2}\nonumber
|\sum_{\lfloor\frac{u}{2}\rfloor<n\leq \lfloor\frac{u+t}{2}\rfloor}\chi_N(2n-1+k_1)\chi_N(2n-1+k_2)|&=|\sum_{\lfloor\frac{u}{2}\rfloor<n\leq \lfloor\frac{u+t}{2}\rfloor}\chi_N(n+2^{-1}(k_1-1))\chi_N(n+2^{-1}(k_2-1))|\\
&\ll max\{d, \sqrt{N/d}\;\log{(N/d)}\}
\end{align}
Plug the result from expressions (\ref{alter general 1}) and (\ref{alter general 2}) into (\ref{alter general}), then we get the result we want to prove.
\qed
\vsp
\newline

Based on the result of property \ref{f(x) of order 2}, we will give an estimate of the upper bound of the $P_{v}(i)$, where the sequence $v$ is as defined in Definition \ref{phi}.
\begin{lem}\label{periodic for phi}
Suppose $N=p_1p_2\ldots p_r$, where $p_1<p_2<\dots <p_r$ are distinct odd primes and $r\geq 2$ is finite. Let $\omega$ be the function as defined in Definition \ref{distinct prime}. For each $1\leq i\leq N-1$, denote $i_N=(i,N)$. Then for the sequence $v$ as defined in Definition \ref{phi}, given condition (\ref{epsilon and p_1}) holds, we have
\begin{equation}
|P_{v}(i)|\ll \left\{\begin{array}{cl}
\sqrt{\frac{N}{p_1p_2}}\log{(\frac{N}{p_1p_2})} &,\; if \ \ i_N=1\\[1.5ex]
\frac{i_N}{p_1}&,\;if\ \ \omega(i_N)=r-1\\[1.5ex]
max\{i_N, \sqrt{\frac{N}{i_N}}\log{(\frac{N}{i_N})}\} &,\;otherwise
\end{array}
\right.
\end{equation}
\end{lem}
\pf
For any $1\leq i\leq N-1$, $P_{v}(i)=\sum_{j=0}^{N-1}v_jv_{j+i}$, while from the definition
\[
v_jv_{j+i}\neq 0\Leftrightarrow (j,N)=m_1>1, \ {\text{ and }}\ (j+i,N)=m_2>1.
\]
Suppose $v_jv_{j+i}\neq 0$, put $(m_1,m_2)=d_1$, then $m_1=d_1d_2$, $m_2=d_1d_3$, and $d_1d_2d_3|N$. Write $j=kd_1d_2$, $j+i=sd_1d_3$, then
\begin{equation}\label{eq:relations between factors}
kd_1d_2+i\equiv sd_1d_3\pmod{N},\;\;\;\;{\text{and }}\ (k,\frac{N}{d_1d_2})=(s,\frac{N}{d_1d_3})=1
\end{equation}
Actually, starting with equality (\ref{eq:relations between factors}), we could obtain a series of equalities as following
\begin{align}\nonumber\label{series}
(k+d_3)d_1d_2+i&\equiv(s+d_2)d_1d_3\pmod{N}\\\nonumber
(k+2d_3)d_1d_2+i&\equiv(s+2d_2)d_1d_3\pmod{N}\\
&\cdot\\\nonumber
&\cdot\\\nonumber
(k+Md_3)d_1d_2+i&\equiv(s+Md_2)d_1d_3\pmod{N}\nonumber
\end{align}
where $M=\frac{N}{d_1d_2d_3}-1$.
\vsp
\newline
Denote $(k+nd_3,N/d_1d_2)=g_1$, and $(s+nd_2,N/d_1d_3)=g_2$. Then all of the equalities above give us the following partial sum in $P_{v}$.
\begin{align}\label{seperate into two parts}\nonumber
&\;\sum_{n=0}^{M}v((k+nd_3)d_1d_2)v((s+nd_2)d_1d_3)\\
=&\sum_{\stackrel{n=0}{g_1g_2>1}}^{M}v((k+nd_3)d_1d_2)v((s+nd_2)d_1d_3)+\sum_{\stackrel{n=0}{g_1=g_2=1}}^M\zeta_n\chi_{\frac{N}{d_1d_2}}(k+nd_3)\chi_{\frac{N}{d_1d_3}}(s+nd_2)\\\nonumber
&{\text{by definition. Where }}\zeta_n=\pm 1 {\text{ depending on the }} n {\text{ values }}\\\nonumber
&\sum_{\stackrel{n=0}{g_1=g_2=1}}^M\zeta_n\chi_{\frac{N}{d_1d_2}}(k+nd_3)\chi_{\frac{N}{d_1d_3}}(s+nd_2)\\\nonumber
&=\sum_{n=0}^{M}\zeta_n\chi_{\frac{N}{d_1d_2d_3}}(k+nd_3)\chi_{d_3}(k+nd_3)\chi_{\frac{N}{d_1d_2d_3}}(s+nd_2)\chi_{d_2}(s+nd_2)\\\nonumber
&=\sum_{n=0}^{M}\zeta_n\chi_{\frac{N}{d_1d_2d_3}}(k+nd_3)\chi_{\frac{N}{d_1d_2d_3}}(s+nd_2)\chi_{d_3}(k)\chi_{d_2}(s)\\\nonumber
&=\chi_{d_3}(k)\chi_{d_2}(s)\chi_{\frac{N}{d_1d_2d_3}}(d_3)\chi_{\frac{N}{d_1d_2d_3}}(d_2)\sum_{n=0}^{M}\zeta_n\chi_{\frac{N}{d_1d_2d_3}}(kd_3^{-1}+n)\chi_{\frac{N}{d_1d_2d_3}}(sd_2^{-1}+n)
\end{align}
where $d_2d_2^{-1}\equiv d_3d_3^{-1}\equiv 1({\text{ mod }}\frac{N}{d_1d_2d_3})$. So we have
\begin{equation}\label{partial sum}
|\sum_{\stackrel{n=0}{g_1=g_2=1}}^M\zeta_n\chi_{\frac{N}{d_1d_2}}(k+nd_3)\chi_{\frac{N}{d_1d_3}}(s+nd_2)|=|\sum_{n=0}^{M}\zeta_n\chi_{\frac{N}{d_1d_2d_3}}(kd_3^{-1}+n)\chi_{\frac{N}{d_1d_2d_3}}(sd_2^{-1}+n)|
\end{equation}

Now we take a closer look at the $\zeta_n$ values. From the Definition \ref{phi},
\begin{enumerate}
  \item $\zeta_n=1$, if $\frac{N}{d_1d_2}\equiv\frac{N}{d_1d_3}\equiv N \pmod{4}$
  \item $\zeta_n=(-1)^{(k_n+s_n)}$, if $\frac{N}{d_1d_2}\equiv\frac{N}{d_1d_3}\not\equiv N \pmod{4}$
  \item $\zeta_n=(-1)^{k_n}$, $\frac{N}{d_1d_2}\not\equiv N \pmod{4}$, $\frac{N}{d_1d_3}\equiv N \pmod{4}$.
  \item $\zeta_n=(-1)^{s_n}$, $\frac{N}{d_1d_3}\not\equiv N \pmod{4}$, $\frac{N}{d_1d_2}\equiv N\pmod{4}$.
\end{enumerate}
where $k_n\equiv k+nd_3\pmod{\frac{N}{d_1d_2}}$, $s_n\equiv s+nd_2 \pmod{\frac{N}{d_1d_3}}$.
\newline
For case $1$,
\begin{align}\label{case 1}
&|\sum_{n=0}^{M}\zeta_n\chi_{\frac{N}{d_1d_2d_3}}(kd_3^{-1}+n)\chi_{\frac{N}{d_1d_2d_3}}(sd_2^{-1}+n)|\\\nonumber
=\;\;&|\sum_{n=0}^{M}\chi_{\frac{N}{d_1d_2d_3}}(kd_3^{-1}+n)\chi_{\frac{N}{d_1d_2d_3}}(sd_2^{-1}+n)|\\\nonumber
=\;\;&|P_{\chi_{\frac{N}{d_1d_2d_3}}}(sd_2^{-1}-kd_3^{-1})|
\end{align}
For case $2$, suppose $n_1$ is the first number that $k+n_1d_3\geq \frac{N}{d_1d_2}$, $n_2$ is the first number that $s+n_2d_2\geq \frac{N}{d_1d_3}$. If $n_1=n_2$, then we still have
\[
|\sum_{n=0}^{M}\zeta_n\chi_{\frac{N}{d_1d_2d_3}}(kd_3^{-1}+n)\chi_{\frac{N}{d_1d_2d_3}}(sd_2^{-1}+n)|=|P_{\chi_{\frac{N}{d_1d_2d_3}}}(sd_2^{-1}-kd_3^{-1})|
\]
So suppose $n_1\neq n_2$. Without loss, suppose $n_1<n_2$, noting that all of $d_2$, $d_3$, $\frac{N}{d_1d_2}$ and $\frac{N}{d_1d_3}$ are odd, then we have
\begin{align}\label{case 2}\nonumber
&|\sum_{n=0}^{M}\zeta_n\chi_{\frac{N}{d_1d_2d_3}}(kd_3^{-1}+n)\chi_{\frac{N}{d_1d_2d_3}}(sd_2^{-1}+n)|\\
\leq\;\;&|P_{\chi_{\frac{N}{d_1d_2d_3}}}(sd_2^{-1}-kd_3^{-1})|+2|\sum_{n_1-1<n\leq n_2-1}\chi_{\frac{N}{d_1d_2d_3}}(kd_3^{-1}+n)\chi_{\frac{N}{d_1d_2d_3}}(sd_2^{-1}+n)|
\end{align}

Since cases $3$ are $4$ are similar, let's just  consider case $3$. Then $\zeta_n=(-1)^{k_n}$. Again we let $n_1$ be the first number that $k+n_1d_3\geq \frac{N}{d_1d_2}$, then we have
\begin{align}\label{case 3}\nonumber
&|\sum_{n=0}^{M}\zeta_n\chi_{\frac{N}{d_1d_2d_3}}(kd_3^{-1}+n)\chi_{\frac{N}{d_1d_2d_3}}(sd_2^{-1}+n)|\\\nonumber
=\;\;&|\sum_{0<n\leq n_1-1}(-1)^n\chi_{\frac{N}{d_1d_2d_3}}(n+kd_3^{-1})\chi_{\frac{N}{d_1d_2d_3}}(n+sd_2^{-1})\\\nonumber
-\;\;&\sum_{n_1-1<n\leq M}(-1)^n\chi_{\frac{N}{d_1d_2d_3}}(n+kd_3^{-1})\chi_{\frac{N}{d_1d_2d_3}}(n+sd_2^{-1})|\\\nonumber
&\leq |\sum_{0<n\leq n_1-1}(-1)^n\chi_{\frac{N}{d_1d_2d_3}}(n+kd_3^{-1})\chi_{\frac{N}{d_1d_2d_3}}(n+sd_2^{-1})|\\
+\;\;&|\sum_{n_1-1<n\leq M}(-1)^n\chi_{\frac{N}{d_1d_2d_3}}(n+kd_3^{-1})\chi_{\frac{N}{d_1d_2d_3}}(n+sd_2^{-1})|
\end{align}
If $\frac{N}{d_1d_2d_3}=1$, then all of the expressions (\ref{case 1}), (\ref{case 2}) and (\ref{case 3}) are $o(1)$. So suppose $\frac{N}{d_1d_2d_3}>1$.
\vsp
\newline
\vsp
Put $(i,\frac{N}{d_1d_2d_3})=d$, and $(sd_2^{-1}-kd_3^{-1},\frac{N}{d_1d_2d_3})=d'$, then we will prove that $d=d'$.
\newline
\vsp
Because $d_2d_2^{-1}\equiv 1\pmod{\frac{N}{d_1d_2d_3}}$, and $d_3d_3^{-1}\equiv 1\pmod{\frac{N}{d_1d_2d_3}}$, we suppose $d_2d_2^{-1}=k_1\frac{N}{d_1d_2d_3}+1$, and $d_3d_3^{-1}=k_2\frac{N}{d_1d_2d_3}+1$ for some integers $k_1$ and $k_2$, then from (\ref{eq:relations between factors}), we have
\begin{align*}
(sd_2^{-1}-kd_3^{-1})d_1d_2d_3&=sd_1d_3(k_1\frac{N}{d_1d_2d_3}+1)-kd_1d_2(k_2\frac{N}{d_1d_2d_3}+1)\\\nonumber
&=\frac{N}{d_1d_2d_3}(sk_1d_1d_3-kk_2d_1d_2)+(sd_1d_3-kd_1d_2)\\
&\equiv\frac{N}{d_1d_2d_3}(sk_1d_1d_3-kk_2d_1d_2)+i\;\pmod{N}
\end{align*}
\vsp
$(i,\frac{N}{d_1d_2d_3})=d\Rightarrow d|\frac{N}{d_1d_2d_3}(sk_1d_1d_3-kk_2d_1d_2)+i
\Rightarrow d|(sd_2^{-1}-kd_3^{-1})d_1d_2d_3$,  but $(d,d_1d_2d_3)=1\Rightarrow d|(sd_2^{-1}-kd_3^{-1})$,
and $d|\frac{N}{d_1d_2d_3}$, therefore, $d|d'$.
\vsp
\newline
On the other hand, $d'|sd_2^{-1}-kd_3^{-1}\Rightarrow d'|\frac{N}{d_1d_2d_3}(sk_1d_1d_3-kk_2d_1d_2)+i$,
\vsp
and $d'|\frac{N}{d_1d_2d_3}\Rightarrow d'|i$, therefore $d'|d$.
\newline
\vsp
So now we have $(i,\frac{N}{d_1d_2d_3})=(sd_2^{-1}-kd_3^{-1},\frac{N}{d_1d_2d_3})$.
\newline
Put $(i,\frac{N}{d_1d_2d_3})=i_D\Rightarrow (sd_2^{-1}-kd_3^{-1},\frac{N}{d_1d_2d_3})=i_D$.
\vsp
By lemma \ref{chi} and \ref{periodic correlation of sequence product},  expressions $(\ref{case 1})$  satisfies
\begin{equation}\label{perio}
|P_{\chi_{\frac{N}{d_1d_2d_3}}}(sd_2^{-1}-kd_3^{-1})|\leq i_D
\end{equation}
From Property \ref{f(x) of order 2}, equations $(\ref{case 2})$ and $(\ref{case 3})$ satisfy
\begin{equation}\label{alter}
|\sum_{n=0}^{M}\zeta_n\chi_{\frac{N}{d_1d_2d_3}}(kd_3^{-1}+n)\chi_{\frac{N}{d_1d_2d_3}}(sd_2^{-1}+n)|\ll max\{i_D, \sqrt{\frac{N}{d_1d_2d_3i_D}}\log{(\frac{N}{d_1d_2d_3i_D})}\}
\end{equation}
where $\omega=\omega(\frac{N}{d_1d_2d_3i_D})$.
\vsp
\newline
If $i_N=1$, then $i_D=1$, we want to show that $\omega( d_1d_2d_3)\geq 2$. Because if $\omega( d_1d_2d_3)=1$, then $d_1=p_j$ for some $1\leq j\leq r$, $d_2=d_3=1$. Then expression (\ref{eq:relations between factors}) becomes
\[
kp_j+i\equiv sp_j\pmod{N}\;\Rightarrow\;\;p_j|\;i\Rightarrow\;\;i_N\geq p_j
\]
which contradicts to the hypothesis that $i_N=1$.
\newline
If $i_N=i_D=1$, and $\omega( d_1d_2d_3)\geq 2$, then expression (\ref{alter}) satisfies
\[
|\sum_{n=0}^{M}\zeta_n\chi_{\frac{N}{d_1d_2d_3}}(kd_3^{-1}+n)\chi_{\frac{N}{d_1d_2d_3}}(sd_2^{-1}+n)|\leq 2^{r-2}\sqrt{\frac{N}{p_1p_2}}\log{(\frac{N}{p_1p_2})}
\]
And $i_N=1$ implies expression (\ref{perio})
\[
|P_{\chi_{\frac{N}{d_1d_2d_3}}}(sd_2^{-1}-kd_3^{-1})|=1.
\]
So we have proved that when $i_N=1$,
\[
|\sum_{\stackrel{n=0}{g_1=g_2=1}}^{M}\zeta_n\chi_{\frac{N}{d_1d_2}}(k+nd_3)\chi_{\frac{N}{d_1d_3}}(s+nd_2)|<1+2^{r-1}\sqrt{\frac{N}{p_1p_2}}\log{(\frac{N}{p_1p_2})}
\]
Next we want to show that $\omega(i_D)\leq r-2$. If $\omega( i_D) = r-1$, then $i_D=N/p_j$, for some $1\leq j\leq r$, $d_1=p_j$ and $d_2=d_3=1$. Then equation (\ref{eq:relations between factors}) becomes
\[
kp_j+i\equiv sp_j\pmod{N}\;\Rightarrow\;\;p_j|\;i\Rightarrow\;\;N|\;i
\]
it contradicts to the hypothesis that $i<N$. Thus $\omega( i_D) \leq r-2$.
\vsp
\newline
Now suppose $\omega(i_N) = r-1$, then from the above statement we have just proved,
\[
i_D\leq i_N/p_1
\]
use the similar argument to before, expression (\ref{perio})
\[|P_{\chi_{\frac{N}{d_1d_2d_3}}}(sd_2^{-1}-kd_3^{-1})|\leq i_D\leq i_N/p_1,
\]
while expression (\ref{alter})
\[
|\sum_{n=0}^{M}\zeta_n\chi_{\frac{N}{d_1d_2d_3}}(kd_3^{-1}+n)\chi_{\frac{N}{d_1d_2d_3}}(sd_2^{-1}+n)|\ll max\{i_D, \sqrt{\frac{N}{d_1d_2d_3i_D}}\log{(\frac{N}{d_1d_2d_3i_D})}\}\ll i_N/p_1
\]
where $\omega=\omega(\frac{N}{d_1d_2d_3i_D})$.
\vsp
\newline
Finally, if $1\leq \omega( i_N)\leq r-2$, then $i_D\leq i_N$, so equation
\[
|P_{\chi_{\frac{N}{d_1d_2d_3}}}(sd_2^{-1}-kd_3^{-1})|\leq i_D\leq i_N,
\]
equation (\ref{alter})
\begin{align*}
|\sum_{n=0}^{M}\zeta_n\chi_{\frac{N}{d_1d_2d_3}}(kd_3^{-1}+n)\chi_{\frac{N}{d_1d_2d_3}}(sd_2^{-1}+n)|&\ll max\{i_D, \sqrt{\frac{N}{d_1d_2d_3i_D}}\log{(\frac{N}{d_1d_2d_3i_D})}\}\\
&\ll max\{i_N, \sqrt{\frac{N}{i_N}}\log{(\frac{N}{i_N})}\}
\end{align*}

Now for the rest term of expression (\ref{seperate into two parts}).
\[
\sum_{\stackrel{n=0}{g_1g_2>1}}^{M}v((k+nd_3)d_1d_2)\cdot v((s+nd_2)d_1d_3)\neq 0
\]
It means that there exist another set of factors $d_1'$, $d_2'$ and $d_3'$, with $d_1'd_2'd_3'|N$ such that
\[
k'd_1'd_2'+i\equiv s'd_1'd_3'\;\pmod{N}
\]
where $(k',\frac{N}{d_1'd_2'})=(s',\frac{N}{d_1'd_3'})=1$. Then we can set up another series of equalities similar to (\ref{series}) and obtain the same upper bound as before. Keep doing this, we could come up with the following
\[
|P_{v}(i)|\leq \sum_{1<d|N}|\sum_{n=0}^{M}\zeta_n\chi_{\frac{N}{d}}(n+k_{d})\chi_{\frac{N}{d}}(n+s_{d})|
\]
where $\zeta_n=\{+1, -1\}$ depending on $n$ values, $k_{d}$ and $s_{d}$ are some integers depending on the values of $d$ with $(s_{d}k_{d},\frac{N}{d})=1$ and $(s_{d}-k_{d},\frac{N}{d})=(i,\frac{N}{d})$.
 \vsp
 Noting that
\[
d(N)=\sum_{d|N}1
\]
is a finite number only depending on $r$ value, by the discussion above, we have proved the lemma.
\qed
\vsp
\newline
Now we can prove that $\sum_{i=1}^{N-1}P^2_v(i)\sim O(\frac{N^2}{p_1})$.
\begin{lem}\label{upper bound for pv}
Suppose $N=p_1p_2\dots p_r$, where $p_1<p_2<\dots <p_r$'s are distinct odd primes and $r$ is finite.  Let $v$ be the binary sequences of length $N$ as defined in Definition (\ref{phi}), then
\[
\sum^{N-1}_{i=1}P^2_v(i)\ll N^2/p_1
\]
\end{lem}
\pf
Let $i_N=(i,N)$, then we have
\[
\sum_{i=1}^{N-1}P^2_v(i)=\sum_{i_N=1}P^2_v(i)+\sum_{\omega( i_N)=r-1}P^2_v(i)+\sum_{1\leq\omega( i_N)\leq r-2}P^2_v(i)
\]
From Lemma \ref{periodic for phi},
\begin{equation}\label{3.0}
\sum_{i_N=1}P^2_v(i)\ll N\times \frac{N}{p_1p_2}\;\log^2{(\frac{N}{p_1p_2})}\ll \frac{N^2}{p_1}
\end{equation}
\begin{equation}\label{3.1}
\sum_{\omega( i_N)=r-1}P^2_v(i)=\sum_{j=1}^{r}\sum_{s=1}^{p_j-1}P^2_v(s\frac{N}{p_j})\ll \sum_{j=1}^{r} \frac{N^2}{p_1^2p_j}\ll \frac{N^2}{p_1}
\end{equation}
where $\omega$ is the function as in Definition \ref{distinct prime}.
\newline
Note that
\begin{align}\label{3.2}
\sum_{1\leq\omega( i_N)\leq r-2}P^2_v(i)&=\sum_{\stackrel{d|N}{1\leq\omega( d)\leq r-2}}\sum_{m=1}^{N/d}\mathtt{'}P_v^2(md)\\\nonumber
&\ll \sum_{\stackrel{d|N}{1\leq\omega( d)\leq r-2}}\sum_{m=1}^{N/d}\mathtt{'} ( max \{d,\sqrt{N/d}\; \log{(N/d)}\})^2\\
&\leq\sum_{\stackrel{d|N}{1\leq\omega( d)\leq r-2}}\frac{N}{d}\cdot ( max \{d,\sqrt{N/d}\; \log{(N/d)}\})^2\ll \frac{N^2}{p_1}
\end{align}
The last inequality follows from the fact that $r$ is finite.
\vsp
\newline
Combine the results from equations (\ref{3.0}), (\ref{3.1}) and (\ref{3.2}), we have
\[
\sum_{i=1}^{N-1}P^2_v(i)\ll \frac{N^2}{p_1}
\]
\qed
\vsp
\newline
Now we have proved that both $\sum_{i=1}^{N-1}P_U^2(i)$ and $\sum_{i=1}^{N-1}P_v^2(i)$ are $\sim O(\frac{N^2}{p_1})$. In next subsection, we will prove that $\sum_{i=1}^{N-1}P_z^2(i)\sim O(\frac{N^2}{p_1})$.
\subsection{Upper Bound for $\sum_{i=1}^{N-1}P_z^2(i)$}\label{section 3.3}
We still need one more property before we could prove that $\sum_{i=1}^{N-1}P_z^2(i)\sim O(\frac{N^2}{p_1})$.
\begin{pro}\label{common factor}
For any positive integers $i$, $m$, and $d$, if $N=md$, then
\[
d\cdot (i,\frac{N}{d})=d\cdot (i,m)\geq (i,N)
\]
\end{pro}
\pf
The result is obvious because
\[
(i,N)=(i,md)=(i,m)\cdot (i,d) \text{ and }d\geq (i,d)
\]
\qed
\begin{lem}\label{estimate for periodic}
Suppose $N=p_1p_2\dots p_r$, where $p_1<p_2<\dots <p_r$'s are distinct odd primes and $r$ is finite.  Let $z$ be the binary sequences of length $N$ as defined in Definition (\ref{phi}), then
\[
\sum^{N-1}_{i=1}P^2_z(i)\ll N^2/p_1
\]
\end{lem}
\pf
Let the binary sequence $U$ be as defined in form (\ref{base sequence}), then $z_j=U_j+v_j$, where sequence $v$ is as defined in Definition \ref{phi}. So
\begin{align}\label{periodic form for p_1p_r}\nonumber
\sum_{i=1}^{N-1}P^2_z(i)
&=\sum^{N-1}_{i=1}(\sum_{j=0}^{N-1}z_jz_{j+i})^2=\sum^{N-1}_{i=1}[\sum_{j=0}^{N-1}(U_j+v_j)(U_{j+i}+v_{j+i})]^2\\\nonumber
&=\sum^{N-1}_{i=1}[P_U(i)+P_{U,v}(i)+P_{v,U}(i)+P_v(i)]^2\\\nonumber
&=\sum^{N-1}_{i=1}P^2_U(i)+\sum^{N-1}_{i=1}P^2_v(i)\\\nonumber
&+\sum^{N-1}_{i=1}[2P_U(i)P_{U,v}(i)+2P_U(i)P_{v,U}(i)+2P_U(i)P_v(i)]\\\nonumber
&+\sum^{N-1}_{i=1}[2P_v(i)P_{v,U}(i)+2P_v(i)P_{U,v}(i)]\\\nonumber
&+\sum^{N-1}_{i=1}[2P_{U,v}(i)P_{v,U}(i)+P_{U,v}^2(i)+P_{v,U}^2(i)]\\
&=A+B+C+D+E
\end{align}
In expression (\ref{periodic form for p_1p_r}), we have separated the summands into five groups. For instance, $B=\sum^{N-1}_{i=1}P^2_v(i)$, and $E=\sum^{N-1}_{i=1}[2P_{U,v}(i)P_{v,U}(i)+P_{U,v}^2(i)+P_{v,U}^2(i)]$. In the following, we will show that the
absolute value of every sum from the same group  has the same upper bound. To simplify the notation, it should be understood that all of the following statements are valid when  $p_1$ and $p_2$'s are \textbf{large enough}.
\newline
For group A, from Lemma \ref{P for product}
\[
\sum^{N-1}_{i=1}P^2_U(i)\ll N^2/p_1
\]
For group B, by Lemma \ref{upper bound for pv}, we have
\[
\sum^{N-1}_{i=1}P^2_v(i)\ll N^2/p_1
\]
For group C, every term in this group could be written as
\[
\sum_{i=1}^{N-1}P_U(i)\sum_{m=0}^{N-1}v_m\xi_m,\;\;\;\;\;\text{where}\;\;\;\;\;\xi_m\in\{+1,-1\}.
\]
Lemma \ref{chi}, \ref{Euler}, \ref{periodic correlation of sequence product} and Lemma \ref{P for product} give
\begin{align}\nonumber
|\sum^{N-1}_{i=1}P_U(i)\sum_{m=0}^{N-1}v_m\zeta_m|&\leq rN/p_1\times \sum^{N-1}_{i=1}|P_U(i)|\\\nonumber
&=r N/p_1\times[\;\sum_{i=1}^{N-1}\mathtt{'}|P_U(i)|+\sum_{\stackrel{i=1}{(i,N)>1}}^{N-1}|P_U(i)|\;]\\\nonumber
&< rN/p_1\times [\;N+\sum_{d|N}\sum_{k=1}^{N/d}\mathtt{'}|P_U(kd)|\;]\\\nonumber
&\leq rN/p_1\times [\;N+\sum_{d|N}N/d\times d\;]\\\nonumber
&\ll N^2/p_1
\end{align}
Again, the inequality second to the last follows from the fact that $d(N)$ is a finite number.
\newline
Now we consider the terms in group E. Since sequence $U$ and $v$ have the same symmetric property as shown in Lemma \ref{sym or antisym}, then we have
\begin{align}\label{subscript shift}\nonumber
\sum^{N-1}_{i=1}P_{U,v}^2(i)&=\sum^{N-1}_{i=1}\sum^{N-1}_{j=0}\sum_{m=0}^{N-1}U_jv_{j+i}U_mv_{m+i}\\\nonumber
&=\sum^{N-1}_{i=1}\sum^{N-1}_{j=0}U_jv_{j+i}\sum_{m=0}^{N-1}U_{N-m}v_{N-m-i}\\
&=\sum_{i=1}^{N-1}\sum^{N-1}_{j=0}\sum_{m=0}^{N-1}U_jv_{j+i}v_mU_{m+i}=\sum_{i=1}^{N-1}P_{U,v}(i)P_{v,U}(i)
\end{align}
The third equality follow from Lemma \ref{sym or antisym}. Similarly we can prove that
\[
\sum^{N-1}_{i=1}P_{v,U}^2(i)=\sum_{i=1}^{N-1}P_{U,v}(i)P_{v,U}(i)
\]
Therefore for every term in group E, it is enough to estimate the upper bound of $\sum_{i=1}^{N-1}P_{U,v}(i)P_{v,U}(i)$.
\begin{align}\nonumber
\sum_{i=1}^{N-1}P_{U,v}(i)P_{v,U}(i)&=\sum_{i=1}^{N-1}\sum^{N-1}_{j=0}\sum_{m=0}^{N-1}U_jv_{j+i}v_mU_{m+i}\\\nonumber
&=\sum_{i=1}^{N-1}\sum^{N-1}_{j=0}\sum_{m=0}^{N-1}v_jv_mU_{j-i}U_{m+i}\\\nonumber
&=\chi_N(-1)\sum^{N-1}_{j=0}\sum_{m=0}^{N-1}v_jv_m(\sum_{i=1}^{N-1}U_{j-i}U_{-m-i})\\\nonumber
&=\chi_N(-1)\sum^{N-1}_{j=0}\sum_{m=0}^{N-1}v_jv_mP_U(m+j)\\\nonumber
&\text{while}\\\nonumber
|\sum^{N-1}_{j=0}\sum_{m=0}^{N-1}v_jv_mP_U(m+j)|
&=|\sum^{N-1}_{s=0}\sum_{j=0}^{N-1}v_{j}v_{s-j}P_U(s)|\hspace{2.5cm}\text{ where }s=m+j\\\nonumber
&\leq|\sum^{N-1}_{\stackrel{j=0}{s=0}}v^2_{j}P_U(0)|+|\sum_{\stackrel{s=1}{(s,N)>1}}^{N-1}P_v(s)P_U(s)|+\sum_{s=1}^{N}\mathtt{'}|P_v(s)|\\\nonumber
\end{align}
From Lemma \ref{Euler},
\begin{equation}\label{1.1}
|\sum^{N-1}_{j=0}v^2_{j}P_U(0)|=N\sum^{N-1}_{j=0}v^2_{j}\leq rN\times N/p_1\ll N^2/p_1
\end{equation}
In the proof of Lemma \ref{periodic for phi}, we know that when $(s,N)=1$, $|P_v(s)|\leq 2^{r-2}\sqrt{\frac{N}{p_1p_2}}\;\log{(\frac{N}{p_1p_2})}$, then when $r\geq 2$,
\begin{equation}\label{1.2}
\sum_{s=1}^{N}\mathtt{'}|P_v(s)|\ll N\times \sqrt{\frac{N}{p_1p_2}}\;\log{(\frac{N}{p_1p_2})}\ll N^2/p_1
\end{equation}
We will have to be more careful in estimating $|\sum_{(s,N)>1}^{N-1}P_v(s)P_U(s)|$, we write
\begin{equation}\label{seperate 2}
|\sum_{\stackrel{s=1}{(s,N)>1}}^{N-1}P_v(s)P_U(s)|\leq |\sum_{\omega( (s,N))=r-1}P_v(s)P_U(s)|+|\sum_{1\leq\omega( (s,N))\leq r-2}P_v(s)P_U(s)|
\end{equation}
For the first item of equation (\ref{seperate 2}), and by Lemma \ref{periodic for phi}, we have
\begin{align}\label{2.1}
|\sum_{\omega( (s,N))=r-1}P_v(s)P_U(s)|&=|\sum_{k=1}^{r}\sum_{m=1}^{p_k-1}P_v(mN/p_k)P_U(mN/p_k)|\\\nonumber
&\leq  \sum_{k=1}^{r}\sum_{m=1}^{p_k-1}|P_v(mN/p_k)|\times|P_U(mN/p_k)|\\\nonumber
&\leq \sum_{k=1}^{r}\sum_{m=1}^{p_k-1}\frac{N}{p_1p_k}\times \frac{N}{p_k}\\\nonumber
&< \sum_{k=1}^{r}\frac{N^2}{p_1p_k}\\\nonumber
&\ll \frac{N^2}{p_1}
\end{align}
Now for the second item of equation (\ref{seperate 2}),
\begin{align}\label{2.2}
|\sum_{1\leq\omega( (s,N))\leq r-2}P_v(s)P_U(s)|&=|\sum_{\stackrel{d|N}{1\leq\omega( d)\leq r-2}}\;\;\sum_{m=1}^{N/d}\mathtt{'}P_v(md)P_U(md)|\\\nonumber
&\leq \sum_{\stackrel{d|N}{1\leq\omega( d)\leq r-2}}\;\;\sum_{m=1}^{N/d}\mathtt{'}|P_v(md)|\times|P_U(md)|\\\nonumber
&\leq \sum_{\stackrel{d|N}{1\leq\omega( d)\leq r-2}}\;\;\sum_{m=1}^{N/d}\mathtt{'}max\{d, \sqrt{N/d}\cdot\log{(N/d)}\}\cdot d\\\nonumber
&\leq \sum_{\stackrel{d|N}{1\leq\omega( d)\leq r-2}}\;\;N/d\cdot max\{d, \sqrt{N/d}\cdot\log{(N/d)}\}\cdot d\\\nonumber
&=\sum_{\stackrel{d|N}{1\leq\omega( d)\leq r-2}}\;\;N\cdot max\{d, \sqrt{N/d}\cdot\log{(N/d)}\}\ll \frac{N^2}{p_1}
\end{align}
the last inequality follows from the fact that when $r$ is finite,
\[
d(N)=\sum_{d|N}1\;\; \text{is finite}
\]
\newline
Now equations (\ref{1.1}) and (\ref{1.2}),(\ref{2.1}) and (\ref{2.2}) give us
\[
|\sum_{i=1}^{N-1}P_{U,v}(i)P_{v,U}(i)|=|\sum_{i=1}^{N-1}\sum^{N-1}_{j=0}\sum_{m=0}^{N-1}U_jv_{j+i}v_mU_{m+i}|=|\sum^{N-1}_{j=0}\sum_{m=0}^{N-1}v_jv_mP_U(m+j)|\ll \frac{N^2}{p_1}
\]
Finally, for the items in group D, we firstly consider $\sum^{N-1}_{i=1}P_v(i)P_{v,U}(i)=\sum^{N-1}_{i=1}P_v(i)\sum_{j=0}^{N-1}v_jU_{j+i}$. We will use the similar method to the proof for Lemma \ref{periodic for phi} to give an upper estimate of $\sum_{j=0}^{N-1}v_jU_{j+i}$.
\vsp
\newline
From Lemma \ref{Euler} and \ref{periodic for phi}, we have
\[
v_jU_{j+i}\neq 0 \Leftrightarrow (j,N)=d>1\;{\text{ and }}(j+i,N)=1
\]
Write $j=kd$, $j+i\equiv s\pmod{N}$. So ,  $(k,{N/d})=(s,N)=1$.
\vsp
\newline
Again, we can set up the following series of equalities, noting that all the values are taken modulo $N$.
\begin{align}\nonumber\label{series2}
kd+i&\equiv s\pmod{N}\\\nonumber
(k+1)d+i&\equiv s+d\pmod{N}\\
&\cdot\\\nonumber
&\cdot\\\nonumber
(k+(M-1))d+i&\equiv s+(M-1)d\pmod{N}\nonumber
\end{align}
where $M=\frac{N}{d}$.
\newline
The equation series in (\ref{series2}) give the following partial sum of $\sum_{j=0}^{N-1}v_jU_{j+i}$
\begin{align*}
\sum_{m=0}^{M-1}\zeta_m\chi_{\frac{N}{d}}(m)\chi_{N}(md+i)&=\sum_{m=0}^{M-1}\zeta_m\chi_d(md+i)\chi_{\frac{N}{d}}(m)\chi_{\frac{N}{d}}(md+i)\\
&=\chi_d(i)\chi_{\frac{N}{d}}(d)\sum_{m=0}^{M-1}\zeta_m\chi_{\frac{N}{d}}(m)\chi_{\frac{N}{d}}(m+id^{-1})
\end{align*}
where $\zeta_m=+1$, or $(-1)^{m'}$ with $m'\equiv m\pmod{N/d}$,  $dd^{-1}\equiv 1 \pmod {N/d}$. Thus
\[
|\sum_{m=0}^{M-1}\zeta_m\chi_{\frac{N}{d}}(m)\chi_{N}(md+i)|=|\sum_{m=0}^{M-1}\zeta_m\chi_{\frac{N}{d}}(m)\chi_{\frac{N}{d}}(m+id^{-1})|
\]
If $\zeta_m=+1$, for $0\leq m\leq M-1$, then by Lemma \ref{periodic for phi}, we have
\begin{equation}\label{4.1}
|\sum_{m=0}^{M-1}\zeta_m\chi_{\frac{N}{d}}(m)\chi_{\frac{N}{d}}(m+id^{-1})|=|P_{\chi_{\frac{N}{d}}}(id^{-1})|=(id^{-1},\frac{N}{d})=(i,\frac{N}{d})
\end{equation}
since $(d,\frac{N}{d})=1\Rightarrow (d^{-1},\frac{N}{d})=1\Rightarrow (id^{-1},\frac{N}{d})=(i,\frac{N}{d})$.
\vsp
\newline
If $\zeta_m=(-1)^{m'}$, where $m'\equiv m \pmod {N/d}$, then just repeating the process in expression (\ref{case 3}) and using Lemma \ref{f(x) of order 2}, we can obtain
\[
|\sum_{m=0}^{M-1}(-1)^{m'}\chi_{\frac{N}{d}}(m)\chi_{\frac{N}{d}}(m+id^{-1})|\ll max\{i_{N/d}, \sqrt{\frac{N/d}{i_{N/d}}}\ \log{(\frac{N/d}{i_{N/d}})} \}
\]
where $i_{N/d}=(i,N/d)$, $\omega=\omega(\frac{N/d}{i_{N/d}})$.
\newline
It is obviously true that $i_{N/d}=(i,N/d)\leq (i,N)=i_N$. And from Property \ref{common factor}, we know that $\frac{N/d}{i_{N/d}}\leq \frac{N}{i_N}$, therefore we have
\begin{equation}\label{4.2}
|\sum_{m=0}^{M-1}(-1)^{m'}\chi_{\frac{N}{d}}(m)\chi_{\frac{N}{d}}(m+id^{-1})|\ll max\{i_{N}, \sqrt{\frac{N}{i_{N}}}\; \log{(\frac{N}{i_{N}})} \}
\end{equation}
For the rest items in $\sum_{j=0}^{N-1}v_jU_{j+i}$, we use the similar argument to before. Since $d(i_N)$ is a finite number, we have the following
\begin{equation}\label{5.1}
|\sum_{j=0}^{N-1}v_jU_{j+i}|\ll max\{i_{N}, \sqrt{\frac{N}{i_{N}}}\log{(\frac{N}{i_{N}})} \}
\end{equation}
With Lemma \ref{periodic for phi} and expression (\ref{5.1}), we can give an upper estimate to the items in group D:
\begin{align*}
|\sum^{N-1}_{i=1}P_v(i)(\sum_{j=0}^{N-1}v_jU_{j+i})|&\leq \sum^{N-1}_{i=1}|P_v(i)|\times |\sum_{j=0}^{N-1}v_jU_{j+i}|\\
&=\sum_{(i,N)=1}|P_v(i)|\times |(\sum_{j=0}^{N-1}v_jU_{j+i})|+\sum_{1<d|N}\sum_{s=1}^{N/d}\mathtt{'}|P_v(sd)|\times |(\sum_{j=0}^{N-1}v_jU_{j+sd})|\\
&=\sum_{(i,N)=1}|P_v(i)|\times |(\sum_{j=0}^{N-1}v_jU_{j+i})|\\
&\;\;+\sum_{1<d|N}\sum_{s=1}^{N/d}\mathtt{'}|P_v(sd)|\times |(\sum_{j=0}^{N-1}v_jU_{j+sd})|\\
&\ll \sum_{(i,N)=1}\frac{N}{\sqrt{p_1p_2}}\; \log^2{(N)}+\sum_{d|N}\sum_{s=1}^{N/d}\mathtt{'} (max \{d, \sqrt{\frac{N}{d}}\log{(\frac{N}{d})}\})^2 \hspace{0.5cm}\text{ by (\ref{5.1})}\\
&\leq \frac{N^2}{\sqrt{p_1p_2}}\;\log^2{(N)}+\sum_{d|N} \frac{N}{d}\cdot (max \{d, \sqrt{\frac{N}{d}}\log{(\frac{N}{d})}\})^2\\
&\ll \frac{N^2}{p_1}
\end{align*}
the last inequality follows from the fact that for each $1\leq k\leq r$, the number of $d|N$ with $\omega( d)=k$ is finite. Using the similar method to expression (\ref{subscript shift}), it can be shown that $P_v(i)P_{v,U}(i)=P_v(i)P_{U,v}(i)$, for any $i=1,\dots,N-1$. Then all of the inequalities above will give us the desired result.
\qed
\vsp
\newline
Now we are ready to prove Theorem \ref{thm2}.
\section{Proof of Theorem \ref{thm2}}
\pf (Theorem \ref{thm2} part(1))
\newline
We denote $\xi_N^j=e^{\frac{2\pi j}{N} i}$. For any sequence $x$ of length $N$, let $x(\xi_N^j)$ be the Discrete Fourier Transform of $x$ as $x(\xi_N^j)=\sum_{k=0}^{N-1}x_k(\xi_N^j)^k$ as in Definition \ref{DFT}. Recall that the interpolation formula (\cite{Hoholdt}, (2.5), page162)
\begin{equation}\label{interpolation}
x(-\xi_N^j)=\frac{2}{N}\sum_{k=0}^{N-1}\frac{\xi_N^k}{\xi_N^k+\xi_N^j}x(\xi_N^k)
\end{equation}
Then for the sequence $z$ as defined in Definition \ref{phi}, we have
\[
z(\xi_N^j)=U(\xi_N^j)+v(\xi_N^j)
\]
Note that from Gauss sum, (for instance, \cite{GT} page233)
\begin{equation}\label{U(+)}
|U(\xi_N^j)|=\left\{\begin{array}{rl}
\sqrt{N},&\ if\ (j,N)=1; \\
0,&\ otherwise
\end{array}
\right.
\end{equation}
Therefore, using the interpolation formula (\ref{interpolation}), we have
\begin{equation}\label{U(-)}
|U(-\xi_N^j)|=|\frac{2}{N}\sum_{k=0}^{N-1}\frac{\xi_N^k}{\xi_N^k+\xi_N^j}U(\xi_N^k)|\leq \frac{2}{\sqrt{N}}\sum_{k=0}^{N-1}|\frac{\xi_N^k}{\xi_N^k+\xi_N^j}|\leq 2\sqrt{N}\log{N}
\end{equation}
Now consider $v(\xi_N^j)$. By definition
\[
v(\xi_N^j)=\sum_{\stackrel{d|N}{d\equiv N(\text{ mod }4)}}\chi_d(\xi_d^j)+\sum_{\stackrel{d|N}{d\not\equiv N(\text{ mod }4)}}\chi_d(-\xi_d^j)
\]
Using the result of Gauss sum
\[
|\sum_{\stackrel{d|N}{d\equiv N(\text{ mod }4)}}\chi_d(\xi_d^j)|\leq \sum_{\stackrel{d|N}{d\equiv N(\text{ mod }4)}}|\chi_d(\xi_d^j)|\ll \sqrt{\frac{N}{p_1}}
\]
Doing similar calculation to (\ref{U(-)}), we have
\[
|\sum_{\stackrel{d|N}{d\not\equiv N(\text{ mod }4)}}\chi_d(-\xi_d^j)|\leq \sum_{\stackrel{d|N}{d\not\equiv N(\text{ mod }4)}}|\chi_d(-\xi_d^j)|\ll \sqrt{\frac{N}{p_1}}\log{(\frac{N}{p_1})}
\]
Then we have obtained
\begin{equation}\label{v(+)}
v(\xi_N^j)\ll \sqrt{\frac{N}{p_1}}\log{(\frac{N}{p_1})}
\end{equation}
Note that
\[
|v(-\xi_N^j)|\leq \sum_{\stackrel{d|N}{d\equiv N(\text{ mod }4)}}|\chi_d(-\xi_d^j)|+\sum_{\stackrel{d|N}{d\not\equiv N(\text{ mod }4)}}|\chi_d(\xi_d^j)|
\]
Then using exactly the same method, we can have
\begin{equation}\label{v(-)}
|v(-\xi_N^j)|\ll \sqrt{\frac{N}{p_1}}\log{(\frac{N}{p_1})}
\end{equation}
Let $\widetilde{F}$ be the merit factor of $U$. Then by Theorem 1.2 of \cite{Borwein} (P35), when condition (\ref{new condition 01}) is satisfied,
\[
\lim_{N\rightarrow \infty}\frac{1}{\widetilde{F}}=\lim_{N\rightarrow \infty}\frac{1}{2N^3}\sum_{j=0}^{N-1}[\;|U[\xi_N^j]|^4+|U[-\xi_N^j]|^4\;]-1=\frac{2}{3}-4|f|+8f^2
\]
where $f=\lfloor\frac{t}{N}\rfloor$ is the offset fraction.
\newline
Now let $F$ be the merit factor of sequence $z$,  then from (\cite{Hoholdt}, (5.4) and (5.7), P624),
\[
1/F=\frac{1}{2N^3}\sum_{j=0}^{N-1}[\;|z(\xi_N^j)|^4+|z(-\xi_N^j)|^4]-1
\]
Let $1/F-1/\widetilde{F}=G/{2N^3}$, Our goal is to prove that the limit of $F$ takes exactly the same form
as $\widetilde{F}$. In other words,
\[
\lim_{N\rightarrow\infty}\frac{1}{F}=\frac{2}{3}-4|f|+8f^2
\]
provided condition (\ref{epsilon and p_1}) is satisfied, where $f=\lfloor\frac{t}{N}\rfloor$ is the offset fraction. So it suffices to prove that
\[
G/{2N^3}\rightarrow0 \;\;\;\text{as }N\rightarrow\infty.
\]
To shorten the notation, put $a_j=v(\xi_N^j)$ and $b_j=v(-\xi_N^j)$, then using the form (\cite{Hoholdt}, (5.10), P624),
\begin{align}\label{G sum}
|G|&\leq \sum_{j=0}^{N-1}[\ |a_j|^4+6|U(\xi_j)|^2|a_j|^2+4(\ |U(\xi_j)|^2+|a_j|^2\ )|a_j|\cdot|U(\xi_j)|\
]\\\nonumber
&+\sum_{j=0}^{N-1}[\ |b_j|^4+6|U(-\xi_j)|^2|b_j|^2+4(\ |U(-\xi_j)|^2+|b_j|^2\ )|b_j|\cdot|U(-\xi_j)|\ ]
\end{align}
Let $|c_j|=\text{max}\{|a_j|,|b_j|\}$, then from expressions (\ref{v(+)}) and (\ref{v(-)}), then
\[
|c_j|\ll \sqrt{\frac{N}{p_1}}\log{(\frac{N}{p_1})}
\]
Let $\mathfrak{U}_j$ be either $U(\xi_j)$ or $U(-\xi_j)$, then by expressions (\ref{U(+)}) and (\ref{U(-)}).
\[
|\mathfrak{U}_j|\ll \sqrt{N}\log{N}
\]
If we apply the results from (\ref{U(+)}), (\ref{U(-)}), (\ref{v(+)}), and (\ref{v(-)}) to (\ref{G sum}), then we could obtain
\begin{align}\label{upper bound of G}
|G|&\leq 2\sum_{j=0}^{N-1}[\ |c_j|^4+6|\mathfrak{U}_j|^2|c_j|^2+4(\ |\mathfrak{U}_j|^2+|c_j|^2\ )|c_j|\cdot|\mathfrak{U}_j|\ ]\\\nonumber
&\ll  \sum_{j=0}^{N-1}[\ |\sqrt{\frac{N}{p_1}}\log{(\frac{N}{p_1})}|^4+6|\sqrt{N}\log{N}|^2|\sqrt{\frac{N}{p_1}}\log{(\frac{N}{p_1})}|^2\\\nonumber
&+4(\ |\sqrt{N}\log{N}|^2+|\sqrt{\frac{N}{p_1}}\log{(\frac{N}{p_1})}|^2\ )|\sqrt{\frac{N}{p_1}}\log{(\frac{N}{p_1})}|\cdot|\sqrt{N}\log{N}|\ ]\\\nonumber
&\leq \sum_{j=0}^{N-1}[\ \frac{N^2}{p_1^2}\log^4{(\frac{N}{p_1})}+6N\log^2{N}\cdot \frac{N}{p_1}\log^2{(\frac{N}{p_1})}\\\nonumber
&+4\cdot \frac{N^2}{\sqrt{p_1}}\log^4{N}+4\cdot \frac{N^2}{p_1^{\frac{3}{2}}}\log^4{N}\ ]\\\nonumber
&\ll \frac{N^3}{\sqrt{p_1}}\log^4{(N)}
\end{align}

Thus given condition (\ref{epsilon and p_1}) is satisfied, we have
\[
\displaystyle\lim_{\small{N\to\infty}}\frac{G}{2N^3}=0
\]
which finishes the proof for part $1$ of Theorem \ref{thm2}.
\qed
\vsp
\newline
Before we could prove part (2) of Theorem \ref{thm2}, we still need the following lemma (\cite{TH}, Lemma 2.7 P933)
\begin{lem}\label{remark}
Suppose $\alpha=\left\{\alpha_0,\alpha_1,....\alpha_{N-1}\right\}$ is a symmetric or antisymmetric binary sequence of odd length $N$. Let the sequence $\epsilon$ of length $2N$ be one of the four sequences $\pm\epsilon^{(\delta)}$ from the definition \ref{defn-seq}. Put $b=\left\{\alpha,\;\alpha\right\}\ast\epsilon$, then
\[
\sum^{2N-1}_{k=1}A_b^2(k)
=N+\sum^{N-1}_{k=1}A_\alpha^2(k)+2\sum^{N-1}_{\stackrel{k=1}
  {\text{even}\, k}}P_\alpha(k)A_\alpha(k)+\sum^{N-1}_{\stackrel{k=1}
  {\text{even}\, k}}P_\alpha(k)^2\,.
\]
\end{lem}
\qed

Now we are ready to prove part (2) of Theorem \ref{thm2}.
\newline
\pf(Theorem \ref{thm2} part(2))
\newline
For $N=p_1p_2\dots p_r$ is odd, lemma \ref{sym or antisym} shows that  sequence $z$ is symmetric or antisymmetric depending the value of $N \pmod{4}$. Thus for
\[
b=\left\{z,\;z\right\}\ast\epsilon
\]
Then lemma \ref{remark} gives
\[
\sum^{2N-1}_{k=1}A_b^2(k)
=N+\sum^{N-1}_{k=1}A_{z}^2(k)+2\sum^{N-1}_{\stackrel{k=1}
  {\text{even}\, k}}P_{z}(k)A_{z}(k)+\sum^{N-1}_{\stackrel{k=1}
  {\text{even}\, k}}P_{z}(k)^2\,.
\]
The proof for part (1) of Theorem \ref{thm2} shows that
\begin{equation}\label{A_{xy}}
2\sum^{N-1}_{k=1}A_{z}^2(k)\sim O(\frac{2}{3}N^2)
\end{equation}
if the condition (\ref{epsilon and p_1}) holds.
Lemma \ref{estimate for periodic} shows that
\[
\sum^{N-1}_{\stackrel{k=1}
  {\text{even}\, k}}P_{z}(k)^2\leq\sum^{N-1}_{k=1}P_{z}(k)^2\ll \frac{N^2}{p_1}
\]
Then given condition (\ref{epsilon and p_1}), by Cauchy-Schwarz inequality
\begin{align*}
\left|\sum^{N-1}_{\stackrel{k=1}{k\ even}}P_{z}(k)A_{z}(k)\right|&\leq
\sqrt{\left[\sum^{N-1}_{\stackrel{k=1}{k\ even}}A_{z}^2(k)\right]\left[\sum^{N-1}_{\stackrel{k=1}{k\ even}}P_{z}^2(k)\right]}\\
&\leq \sqrt{\left[\sum^{N-1}_{k=1}A_{z}^2(k)\right]\left[\sum^{N-1}_{k=1}P_{z}^2(k)\right]}\\
&\ll \frac{N^2}{\sqrt{p_1}}
\end{align*}
Therefore, given condition (\ref{epsilon and p_1}) is hold, the asymptotic merit factor of $b$ is
\begin{align*}
\lim_{N\rightarrow \infty}(F_b)&=\lim_{N\rightarrow \infty}
\frac{(2N)^2}{2(\sum^{2N-1}_{k=1}A_b^2(k))}\\
&=\lim_{N\rightarrow
  \infty}\frac{4N^2}{2\sum^{N-1}_{k=1}A_\alpha^2(k)}\\
&=4\times\frac{3}{2}=6\,.
\end{align*}
This finishes the proof of part(2) of Theorem \ref{thm2}.
\qed
\vsp
\newline
\section{Conclusion}
For a long time, being afraid of losing ideal properties of the real primitive character sequences, people have been passive in changing the values of those $j$-th positions with $(j,N)>1$. However, the authors have shown that we could have more freedom in changing the values on those positions. It could also be possible to construct sequences with asymptotic merit factor exceeding $6.0$ by looking at these subtle positions instead of cyclic shifting the sequence and changing the sequence length. The authors wish that this paper could attract further attention to this new direction.

\medskip

\noindent
\begin{flushright}
{\bf Address:}\\
Department of Mathematics\\
Michigan State University\\
East Lansing, MI 48823, U.S.A.
\end{flushright}
\end{document}